\documentclass[aps,prb,twocolumn,superscriptaddress,longbibliography]{revtex4-2}

\usepackage[T1]{fontenc}
\usepackage[utf8]{inputenc} 

\usepackage{amsmath,amssymb,bm,mathtools}

\usepackage{graphicx}
\usepackage[section]{placeins} 
\usepackage{dcolumn}   
\usepackage{booktabs}  
\usepackage{siunitx}   
\usepackage{xcolor}    
\usepackage{hyperref}  
\hypersetup{
  colorlinks=true,
  linkcolor=blue,
  citecolor=blue,
  urlcolor=black
}

\usepackage{siunitx}

\providecommand{\onecolumngrid}{\onecolumn}

\begin{document}

\title{From Interdependent Networks to Two-Interactions Physical Systems}

\author{Yuval Sallem} 
\thanks{Corresponding author: yuvalsallem15@gmail.com}
\affiliation{Department of Physics, Bar-Ilan University, Ramat-Gan 52900, Israel}
\affiliation{Jack and Pearl Resnick Institute, Bar-Ilan University, Ramat-Gan 52900, Israel}
\affiliation{The Institute of Nanotechnology and Advanced Materials, Bar-Ilan University, Ramat-Gan 52900, Israel}

\author{Nahala Yadid} 
\thanks{Y.S. and N.Y. contributed equally to this work.}
\affiliation{Department of Physics, Bar-Ilan University, Ramat-Gan 52900, Israel}
\affiliation{Jack and Pearl Resnick Institute, Bar-Ilan University, Ramat-Gan 52900, Israel}

\author{Xi Wang} 
\affiliation{Department of Physics, Bar-Ilan University, Ramat-Gan 52900, Israel}
\affiliation{The Institute of Nanotechnology and Advanced Materials, Bar-Ilan University, Ramat-Gan 52900, Israel}

\author{Irina volotsenko} 
\affiliation{Department of Physics, Bar-Ilan University, Ramat-Gan 52900, Israel}
\affiliation{Jack and Pearl Resnick Institute, Bar-Ilan University, Ramat-Gan 52900, Israel}
\affiliation{The Institute of Nanotechnology and Advanced Materials, Bar-Ilan University, Ramat-Gan 52900, Israel}

\author{Bnaya Gross}
\affiliation{Network Science Institute, Northeastern University, Boston, MA 02115}
\affiliation{Department of Physics, Northeastern University, Boston, MA 02115}

\author{Beena Kalisky} 
\affiliation{Department of Physics, Bar-Ilan University, Ramat-Gan 52900, Israel}
\affiliation{The Institute of Nanotechnology and Advanced Materials, Bar-Ilan University, Ramat-Gan 52900, Israel}

\author{Shlomo Havlin}
\affiliation{Department of Physics, Bar-Ilan University, Ramat-Gan 52900, Israel}
\affiliation{Jack and Pearl Resnick Institute, Bar-Ilan University, Ramat-Gan 52900, Israel}

\author{Aviad Frydman}
\thanks{ Corresponding author: aviad.frydman@gmail.com}
\affiliation{Department of Physics, Bar-Ilan University, Ramat-Gan 52900, Israel}
\affiliation{Jack and Pearl Resnick Institute, Bar-Ilan University, Ramat-Gan 52900, Israel}
\affiliation{The Institute of Nanotechnology and Advanced Materials, Bar-Ilan University, Ramat-Gan 52900, Israel}

\begin{abstract}
Recent advances have shown that introducing dependency interactions between two superconducting networks can trigger abrupt, hysteretic normal-superconductor phase transitions. In this study, we demonstrate that such behavior can also arise in a single-network superconducting system that features two distinct types of interactions: short-range electrical connectivity and long-range thermal dependency. Using experimental and simulation methods, we show that when sufficient heat is dissipated within a single-layer disordered superconducting network, the system undergoes a mixed-order phase transition marked by both a discontinuous change in resistance and critical scaling behavior. We find that the emergence and characteristics of these abrupt transitions depend critically on the thermal conductivity of the underlying substrate, establishing heat flow as the origin of the unique phase transition. Additionally, both experimental and numerical results reveal long-lived transient states and scaling dynamics near the critical point, consistent with spontaneous branching processes observed in interdependent networks theory. These findings strongly demonstrate that complex critical phenomena, such as mixed-order transitions, previously attributed to structurally interdependent systems, can also arise within single-layer physical systems when dual interactions coexist. Our results broaden the scope of the theory and experiments of phase transitions in interdependent networks and suggest new ways to design and control phase changes in physical, biological, and technological systems where two interactions are present.

\end{abstract}

\maketitle

Within the framework of conventional statistical physics, second-order phase transitions (PTs) are characterized by a continuous change in the order parameter at the critical point, whereas first-order PTs exhibit a sudden, discontinuous shift between phases \cite{K_Binder_1987}. Unlike first-order PTs, second-order PTs are characterized by critical behavior near the transition point, following well-defined scaling laws \cite{stanley1971phase,domb2000phase}. A distinct class of transitions, known as mixed-order PTs, combines features of both: they exhibit a discontinuous jump in the order parameter, akin to first-order PTs, along with critical scaling and diverging correlation lengths typical of second-order PTs \cite{lee2017universal,mukamel2024mixed,boccaletti2016explosive,gross2022fractal,alert2017mixed}. 

Physical systems that undergo a PT are composed of interacting sub-systems or elements, such as atoms, molecules, grains, spins, or segments in a network \cite{sethna2006statistical,stanley1971phase,domb2000phase,fisher1969phase}. Conventionally, PTs have been primarily attributed to a \textit{single} interaction between elements in a system, leading to continuous second-order transitions, as exemplified by the Ising model for ferromagnetism \cite{onsager1944crystal,kaufman1949crystal,yang1952spontaneous,baxter2011onsager,kaufman1949crystal2}. However, when an additional interaction, such as an external magnetic field, is introduced, the system can exhibit an abrupt first-order transition \cite{cipra1987introduction,glauber1963time}. In the present study, we investigate the impact of two distinct types of interactions on the nature of PTs in superconducting network systems and reveal the underlying mechanism that governs their behavior. 

 A substantial advance in the field of PTs and critical phenomena was achieved by the introduction of \textit{interdependent networks theory} \cite{buldyrev-nature2010}, which demonstrated that coupled networks could behave very differently from single networks due to spontaneous cascades of microscopic changes between the networks that control the system's behavior. The key feature of the theory is the existence of \textit{two types} of links in a system composed of networks. These links represent two qualitatively different kinds of interactions between the network's nodes. Within each network, links between nodes describe {\em connectivity} in the sense that a physical quantity (e.g., electric currents or information) can propagate through each network, moving from one node to another. 
Between networks, on the other hand, links describe {\em dependency} relationships (e.g., magnetic fields or heat dissipation) where a node in one network can function only if the node on which it depends in the other network is also functioning.

The difference between dependency and connectivity links becomes clearer when considering a {\em percolation process} to describe the system's robustness against the failures of its components \cite{gao-prl2011, gao-naturephysics2012}. 
In contrast to a single network, where the connection to the giant connected component can be adopted as a condition for functionality, in the presence of dependency, the functionality of a node is more strict: even if a node is connected to the giant component, it will cease to function if the node upon which it depends in the other network ceases to function.
Therefore, while failures of connectivity links spread damage within the networks, dependency links spread malfunctions between the networks.
The interplay between these two types of links in interdependent networks amplifies the propagation of failures, which, in turn, can ignite spontaneous {\em cascades of failures} that could lead to a \textit{sudden collapse} of the system, in contrast to a single network where the breakdown is continuous. 

Recently, we presented the first experimental system that validates interdependent networks theory in real physical systems \cite{bonamassa2023interdependent}. We developed a controlled system of interdependent superconducting networks (ISN) composed of two thermally coupled layers of superconducting networks. This configuration yields a system characterized by two types of interactions between its elements: current flow \textit{within} each network (but not between the networks) and heat dissipation \textit{between} the networks (see Fig. \ref{RT}a). In this system, heat interdependency is exhibited by a mutual, \textit{abrupt} Normal - Superconductor (N-S) transition of the two thermally coupled superconducting networks. This occurs when sufficient electric current is driven through both networks simultaneously ($250\mu$A versus $150\mu$A), creating enough heat to generate the dependency interaction. 

The paradigm of interdependent networks has been extremely successful in understanding various properties of complex systems \cite{Bashan2013, buldyrev-nature2010}, while the experimental realization of the theory has drastically enhanced the recognition of its universality. However, the fabrication of interdependent networks in real solid-state systems, along with their applications and controllability, poses significant experimental challenges. Here, we show, both experimentally and theoretically, that a \textit{single} superconducting network system characterized by \textit{two} types of interactions, connectivity (i.e., electric current) and dependency (i.e., heat dissipation), both within the same network, exhibits behavior similar to that of ISN. The logic behind this notion is that, in ISN systems, heat dissipation is not restricted to propagating solely between the two networks  but can also flow within each of the networks. This suggests that even within a single superconducting network system, if enough heat is dissipated, dependency interactions could be generated, thus giving rise to abrupt PTs. We dub this systems "Two Interactions Superconducting System" (TISS).

Ultimately, our present study offers a generalization of interdependent networks theory. The main conclusion of this work is that the nature of a PT in an isolated \textit{single} network system depends on the presence or absence of a second dependency interaction type between the network elements and its spatial extent, rather than on the existence of an additional network in the system. This has major implications. Unlike interdependent networks, the manifestation of two-interactions in single layer systems is much simpler, easier and broader. They are much more abundant in nature, and their applications can be implemented more widely. For example, they can be relevant to \textit{information overload} (where connectivity is achieved through information flow and dependencies dictated by the overload of certain nodes due to limited tolerance) \cite{Motter2002,Brummitt2012}, \textit{traffic networks} (where connectivity is achieved by roads and dependency is dictated by traffic jams) \cite{Sugiyama2008,Geroliminis2008}, \textit{neural networks} (electric signals in the brain as connectivity and excitation caused by distant neurons as dependency) \cite{Cabana2013,Deco2011}, and perhaps even \textit{regular solids} (where connectivity is the interatomic interactions and dependency is given by long-range phonons) \cite{Griffa2019,Crespi2024}. Thus, introducing and controlling a second type of interaction within a single network system is of significant importance.


\section{Sample fabrication}

In this study, we fabricated two types of samples. For the first type, TISS, we utilized an electrically insulating substrate that conducts heat (either $\mathrm{Si/SiO}$ or glass), on top of which we e-beam evaporated a film of $50$nm-thick amorphous indium-oxide ($\mathrm{InO}$) at a residual $O_2$ pressure of $6-8 ~\mu$Torr. This results in a disordered superconductor with a distribution of local critical temperatures, while the bulk critical temperature is $T_c \approx 3$K. The layer is patterned to form a network consisting of $L \times L$  segments, each having dimensions of $2 \mu$m wide and $10 \mu$m long (see Fig. \ref{RT}b,c). Electrical contacts of $4$nm-thick $\mathrm{Cr}$ topped with $35$nm $\mathrm{Au}$ were fabricated at the edges of the network to enable transport measurements. 

The second sample type, ISN, comprises two identical superconducting networks that are separated by a $100$nm-thick film of aluminum-oxide ($\mathrm{AlO}$), which is a thermally conducting and electrically insulating medium (see Fig. \ref{RT}a) in order to achieve two electrically isolated networks that can exchange heat \cite{bonamassa2023interdependent}.

\section{results}

\subsection{TISS versus ISN - RT curves}

 \begin{figure*}[!htb]
     \centering
     \includegraphics[width=1\linewidth]{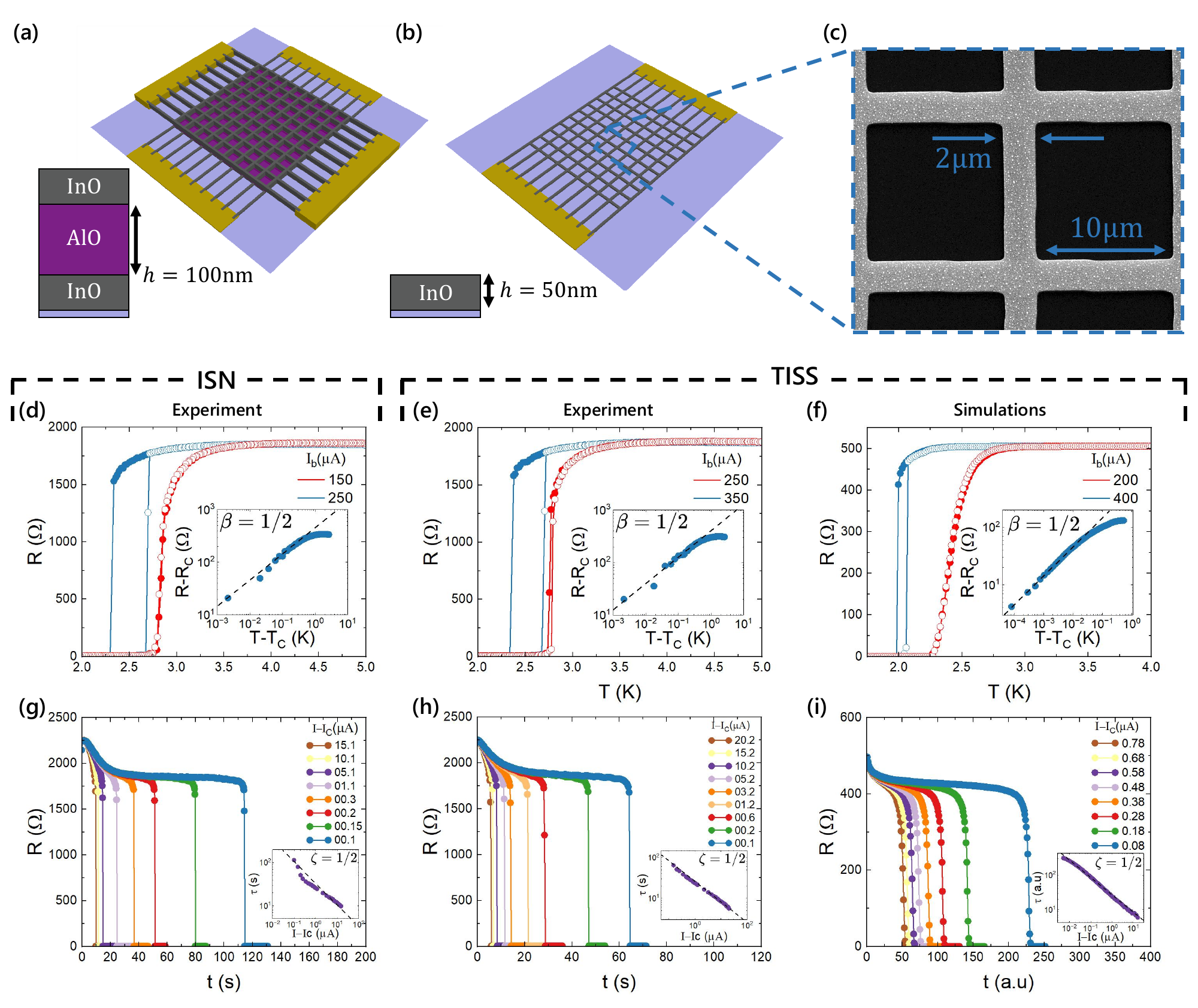}
        \caption{\textbf{TISS versus ISN}
     \textbf{(a)} A sketch of an ISN system. On a thermally conducting and electrically insulating substrate, a network in the shape of a two-dimensional square lattice made from $\mathrm{InO}$ is fabricated. Then, a layer of $\mathrm{AlO}$, which is thermally conducting and electrically insulating, is placed on top of the network. Last, an identical $\mathrm{InO}$ network is fabricated on top of the $\mathrm{AlO}$.
     \textbf{(b)} A sketch of a TISS. A single $\mathrm{InO}$ network is fabricated on a thermally conducting and electrically insulating substrate. 
     \textbf{(c)} A zoom-in side-view image of the superconducting network taken by a scanning electron microscope, showing its geometry. 
     \textbf{(d,e)} Experimental resistance versus temperature measurements at different bias currents for ISN $(L=416)$ and TISS $(L=416)$. Full symbols are for cooling, and empty symbols are for heating. For the ISN sample, electric current is driven to both networks but only the RT curves of one of the networks are presented.
     \textbf{(f)} Corresponding temperature-dependent numerical simulations of a TISS $(L=100)$. Insets show corresponding extracted values of the $\beta$ exponent (Eq. \ref{eq:beta}).
     \textbf{(g,h)} Experimental resistance versus time measurements of the the cascading plateau at various bias currents for ISN $(L=416)$ and TISS $(L=416)$. 
     \textbf{(i)} Corresponding time-dependent numerical simulations of a TISS $(L=100)$, fixed heat-bath temperature $T=2K$, as $I_c = 121.98\mu A$. Insets show corresponding extracted values of the $\zeta$ exponent (Eq. \ref{eq:zeta}).}
     \label{RT}
 \end{figure*}
 
It has been recently shown \cite{bonamassa2023interdependent} that driving a sufficient amount of current, $I_b$, which results in a \textit{continuous} N-S PT in a single network, can generate an \textit{abrupt} PT in ISN. Here, we show that upon applying a higher current, which generates enough heat induced thermal dependency in a \textit{single} superconducting network (i.e., a TISS), an abrupt transition is observed, similar to that in ISN.  This is demonstrated in Figs \ref{RT}d,e, which compare the RT curves of ISN and TISS respectively. The similarity is striking. In both cases, it is observed that for low currents (i.e., a single interaction network), the transitions from N to S and vice versa are continuous, with no signs of hysteresis. However, for sufficiently high current values, where the second (thermal dependency) interaction is present, the transitions become abrupt and exhibit clear hysteresis. Clearly, the TISS requires a higher current value than the ISN system to generate dependency, as it contains fewer heat-emitting segments that can dissipate heat compared to the ISN. Similar results (Fig. \ref{RT}f) for TISS were obtained theoretically by numerically solving Kirchhoff's equations, considering the RSJJ model of disordered superconducting networks (see Methods). These results are consistent with the experimental findings. 

The experimental and theoretical results also show that both TISS and ISN samples exhibit behavior typical of mixed-order PTs, i.e., abrupt transitions accompanied by critical scaling. Furthermore, near the transition, the resistance curve follows:

\begin{figure*}[!htb]
    \centering
    \includegraphics[width=1\linewidth]{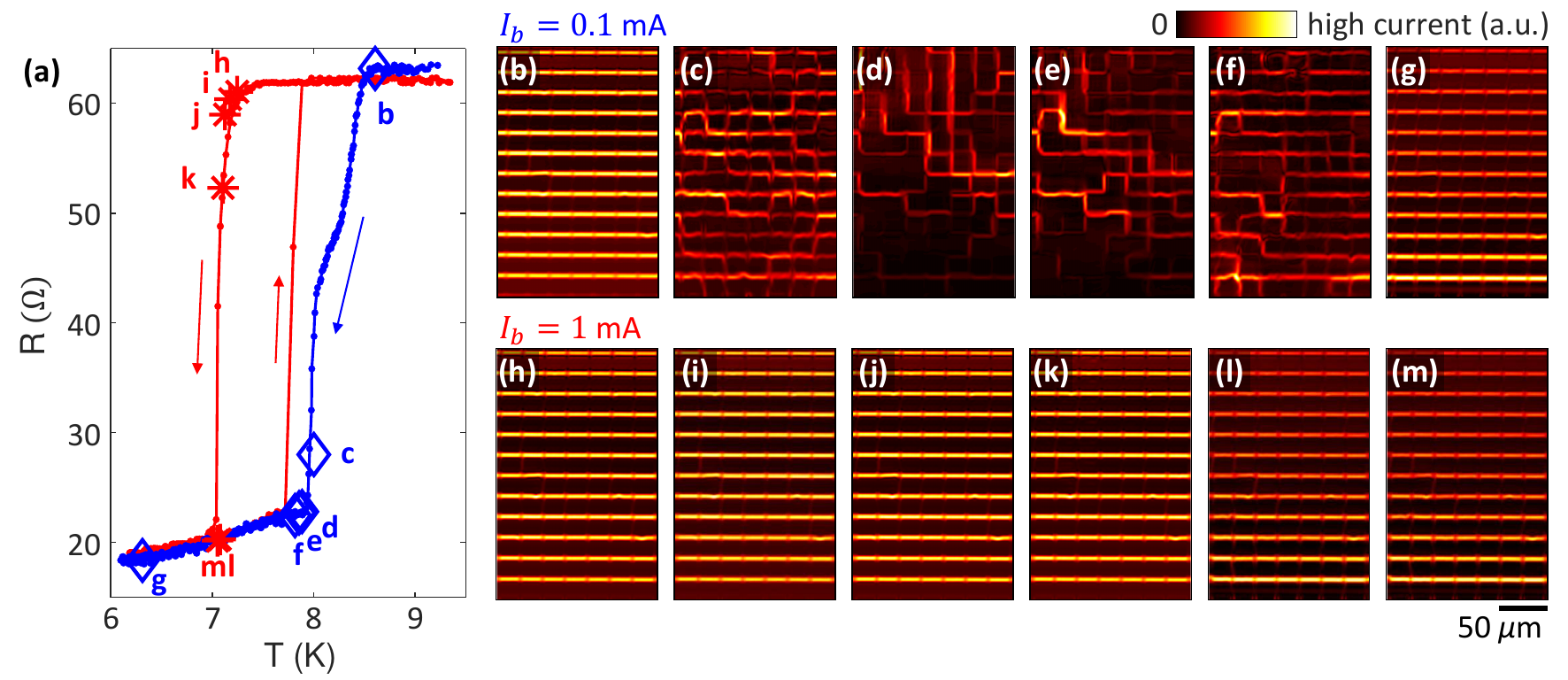}
    \caption{\textbf{Current density by scanning SQUID measurement in Nb network at low and high current regimes}
    \textbf{(a)} Resistance versus temperature curves for $I_b=0.1$mA (blue curve) and $I_b=1$mA (red curve). 
    Symbols mark selected temperature values for which we plot the spatial distribution of the current flow, and the letters refer to the panels.
    \textbf{(b)-(g)} Reconstructed current density maps of a small region of the network at $I_b=0.1$mA, showing a complex structure of currents when transitioning continuously from the N state ($T=8.7$K) to the S state ($T=6.1$K).
    \textbf{(h)-(m)} Corresponding simple structural maps at $I_b=1$mA. The current distribution differs between the superconducting state (panels g, m) and the normal state (panels b, h). In the superconducting state, shielding currents dominate the behavior, concentrating the flow near the network's edge. In the normal state the current distributes evenly between the networks' segments. Numerical results of the current density distributions are shown in the Supplementary Information Figs.~S1 and S2.}
    \label{RT - Squid Imaging}
\end{figure*}

\begin{equation}
    R(T)-R(T_{c})\sim(T - T_{c})^\beta ,
   \label{eq:beta}
\end{equation} 

with $\beta =1/2$ (see insets of Fig. \ref{RT}d-f and Ref. \cite{gross2024microscopic}). This exponent is also found from theoretical derivations for percolation on interdependent networks~\cite{parshani2010} and interdependent ferromagnetic networks \cite{gross2024microscopic2}, thus demonstrating universal behavior.

To visualize the different mechanisms between the continuous and abrupt transitions in our TISS, we performed scanning SQUID (superconducting quantum interference device) measurements \cite{Kirtley1995APL} on a $\mathrm{Nb}$ network \cite{Xi2022PRA}. Here, an Nb network is chosen since the penetration depth of the magnetic field ($\lambda$) in Nb is smaller than that in InO, resulting in a stronger SQUID signal. An AC current was applied to the network, and the resulting magnetic flux was measured with a $1.5$µm diameter pickup loop using a lock-in amplifier. By scanning the SQUID over the network, a field map that describes the current distribution is generated. Using Fourier analysis, we reconstructed  the current density distribution \cite{Roth1989APL}, as shown in Fig. \ref{RT - Squid Imaging}. In transport measurement (Fig. \ref{RT - Squid Imaging}a), we reproduce a continuous transition at low current and an abrupt hysteretic transition at high current, similar to that seen in an InO network. Consistently, the current density maps for low applied bias current (Fig. \ref{RT - Squid Imaging}b-g) show a smooth transition that manifests itself in a complex current-flow percolating network. Whereas, at high current (Fig. \ref{RT - Squid Imaging}h-m), the maps show no complex structure but only an abrupt change between superconducting current and normal current. Numerical simulations produced similar results, which are presented in SI Figs.~S1 and S2.
\begin{figure*}[hbt!]
     \centering
     \includegraphics[width=1\linewidth]{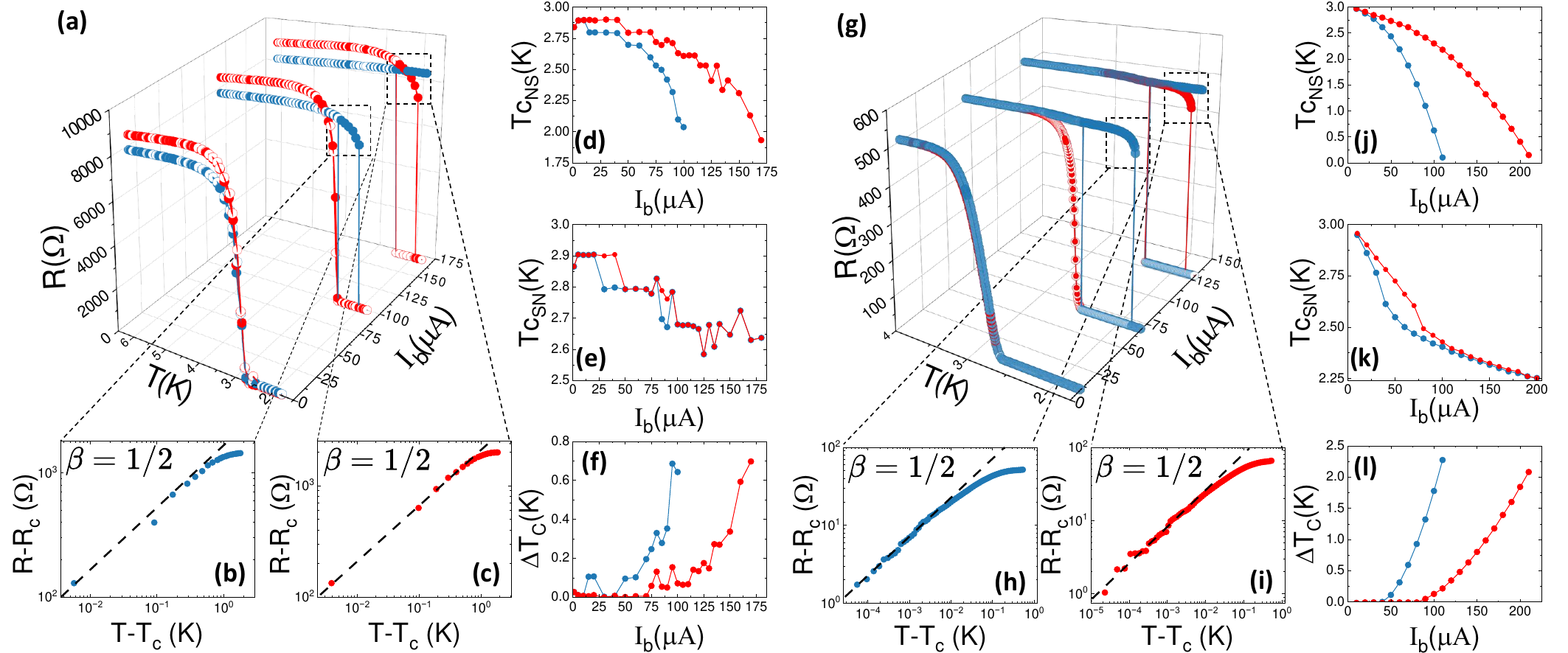}
      \caption{\textbf{Substrate dependence-} 
      \textbf{(a)} Experimental resistance versus temperature measurements for a single network ($L=200$), at different bias currents, fabricated on silicon (red symbols) and glass (blue symbols). Full symbols are for cooling cycles and empty symbols are for heating cycles. 
      \textbf{(b,c)} The extraction of the critical exponent $\beta$ from the cooling curves of glass and silicon at $I_b=100\mu A$ and $I_b=170 \mu A$ respectively. As seen, both yield the same $\beta$. 
      \textbf{(d,e)} Critical temperatures of the transitions from N to S (Cooling) and S to N (heating) plotted versus $I_b$ . 
      \textbf{(f)} The hysteresis width is calculated for each bias current measurement. 
      \textbf{(g)-(l)} Numerical simulations are in excellent agreement with the experimental results ($L=100$). The extraction of the critical exponent $\beta$ from the cooling curves of glass and silicon at $I_b=70\mu A$ and $I_b=130 \mu A$ respectively.}
     \label{substrate}
 \end{figure*}
\subsection{TISS versus ISN - Dynamics}
One of the findings regarding the mixed-order PT demonstrated in ISNs \cite{gross2024microscopic} is that a hidden, novel critical behavior of the microscopic propagation of phase-changes, due to the dependency links, manifests itself in a spontaneous cascade represented by a long-lasting transient "plateau" stage during the abrupt transition, which is characterized by a critical exponent~\cite{dong-pre2014}. This is due to the fact that, at criticality, each system element in one network that changes its phase from N to S or vice versa causes, on average, one new element in the other network to change. During this plateau stage, a microscopic critical {\em branching process} is predicted \cite{dong-pre2014,baxter2015critical}, and the changes in the order parameter are microscopic in nature. However, eventually, after a long plateau that could last for tens or even hundreds of seconds, it leads to a sudden collapse of the system. Indeed, our measurements on ISNs confirm this, as a long plateau of resistance over time is observed (see Fig.\ref{RT}g and \cite{gross2024microscopic}).

Fig.~\ref{RT}h shows resistance versus time measurements of a TISS near criticality. It is seen that even in a single layer, TISS, a long-lived plateau is clearly observed. Furthermore, we find that the plateau duration $\tau$ increases as the current $I$ approaches the critical value $I_c$, according to a scaling relation:
\begin{equation}
    \tau \sim (I_{c} - I_b)^{-\zeta}.
    \label{eq:zeta}
\end{equation}
Indeed, analysis based on percolation on abstract interdependent networks predicts an exponent of $\zeta = 1/2$ \cite{dong-pre2014}, which is also supported by experiments on ISN's (see Fig. \ref{RT}g and Ref. \cite{gross2024microscopic}). Remarkably, the experimental dynamics in a TISS follow the same behavior as seen in Fig. \ref{RT}h. This form of critical slowing down, expressed through long-lived transient plateaus and power-law divergence of $\tau$, aligns with the broader phenomenology described in the theory of dynamic critical phenomena \cite{halperin1977}.


\subsection{Substrate Dependence}

The key hypothesis of this work is that sufficient heat transfer along a single superconducting network can induce dependency interactions that alter the nature of the N-S PT from continuous to abrupt, of mixed-order type. Consequently, the thermal conductivity of the substrate of the network is expected to play a crucial role in this process. 

In our substrate–sample configuration, the geometry dictates that most of the heat generated by a segment in the network is transferred through the thickness of the substrate to the sample holder, which acts as a large heat bath. A smaller portion of the heat propagates laterally along the substrate and the sample, partially raising the local temperature of other network segments and partially escaping to the edges of the substrate. This lateral heat flow depends on the ratio between the thermal conductivities of the substrate and the sample, $\kappa_{\text{sub}}$ and $\kappa_{\text{sam}}$, respectively. Therefore, the lower the value of $\kappa_{\text{sub}}$, the greater the heat retention in the sample region. Hence, somewhat counterintuitively, dependency interactions are expected to be enhanced for substrates with lower thermal conductivity.

To investigate this, we performed resistance versus temperature measurements on two TISS fabricated on different substrates: one on silicon, which has a thermal conductivity of approximately $\kappa \approx 500~\frac{\text{W}}{\text{m}\cdot\text{K}}$ at $T = 2$ K \cite{THOMPSON1961146}, and the other on glass, with $\kappa \approx 0.01~\frac{\text{W}}{\text{m}\cdot\text{K}}$ at the same temperature \cite{PhysRevLett.33.1158}. The results are shown in Fig. \ref{substrate}a. 

At low bias currents ($I_{\text{b}}=1\mu$A), where heat generation is low and insufficient, both networks undergo a continuous second-order PT from N to S and vice versa. As the bias current is increased ($I_{\text{b}}=100\mu$A), the TISS on glass begins to exhibit an abrupt, hysteretic PT, while the TISS on silicon continues to undergo a continuous PT. Further increasing the bias current ($I_{\text{b}}=170\mu$A) causes the TISS on silicon to also display an abrupt, hysteretic PT. Similar results are achieved through simulations, as shown in Fig. \ref{substrate}g. 


Despite the different bias currents needed for the various substrates, both abrupt transitions from N to S are characterized by the same critical behavior and scaling relations corresponding to Eq. \ref{eq:beta} with $\beta =1/2$, as seen in Fig. \ref{substrate}b,c for the experimental results and Fig. \ref{substrate}h,i for the simulations. This indicates an identical mixed-order PT for both TISSs.

The importance and effects of thermal interactions in these superconducting systems can further be appreciated by examining the dependence of the critical temperatures, $T_c$, for both cooling (N-S transition) and heating (S-N transition) processes on the bias current. The relevant experimental and simulation results for the two substrate samples are shown in Fig. \ref{substrate}d,e,j,k. It is observed that for the cooling process, the glass TISS $T_c$s are consistently smaller than those of the silicon TISS for the same $I_b$s, demonstrating the larger heat retention of the glass substrate (Fig. \ref{substrate}d,j). On the contrary, during the heating process, during which the S-N PT occurs before heat diffuses in the system, heat retention is not important. Therefore, the $T_c$s for both networks are not very different (Fig. \ref{substrate}e,k). Similar behavior is observed in experimental resistance versus current curves ($T=$ constant), as shown in Fig. S3.

We note that, despite the quantitative difference in heat transfer, the two samples exhibit similar qualitative behavior, such as a similar hysteretic nature of critical temperatures with similar temperature differences $\Delta T = T_{C\textit{heat}}-T_{C\textit{cool}}$, as seen in Fig. \ref{substrate}f,l. 

\section{Discussion and conclusions}

Our findings demonstrate that a single-layer superconducting network system, subjected to two distinct types of interactions, connectivity via electric current and thermal dependency via heat transfer, displays the abrupt, hysteretic PT previously observed in ISN. This result offers a conceptual shift: complex interdependent behavior does not arise solely in systems composed of structurally distinct networks, but can emerge intrinsically within a single network system if multiple distinct interactions between its elements coexist and couple dynamically. 

In superconducting systems, our results highlight the critical role of thermal diffusion in generating effective dependency interactions within a single superconducting network. We showed that locally generated heat can propagate across the system, triggering cascading phase-changes, akin to a percolation process, thereby transforming a continuous N-S transition into an abrupt mixed-order transition. This mechanism naturally parallels the behavior predicted and observed in theoretical and experimental studies of interdependent networks, where connectivity within each network and dependency links between the networks yield similar critical phenomena.

The dynamics of the abrupt transition further underscore the system's criticality. The emergence of long-lived resistance plateaus near the transition point and their scaling with the bias current reflect a spontaneous microscopic branching process underlying macroscopic state changes, consistent with theoretical predictions of cascading failure in interdependent systems. As in ISNs, the value of the branching factor and its closeness to 1 may be implemented as a warning signal for system collapse \cite{gross2024microscopic}.

Notably, both TISS and ISN exhibit the same critical exponent values (e.g., $\beta=1/2$, $\zeta=1/2$), highlighting a universal scaling behavior (based on two interactions) independent of system geometry or the implementation of interactions \cite{bonamassa2025hybrid}. This universality suggests broader implications. While our system focuses on superconducting networks, the underlying principle of PTs shaped by two interacting mechanisms could be extended to other physical, biological, and technological systems. This insight not only simplifies the realization of interdependence-inspired phenomena in real materials but also broadens the conceptual framework of phase transitions in disordered and complex systems.

Lastly, we demonstrated that the thermal conductivity of the substrate acts as a tunable parameter that directly controls the strength of internal dependency interactions in superconducting network systems. A substrate with low thermal conductivity (e.g., glass) enhances heat retention in the sample region, increasing the likelihood of inter-segment coupling via a local temperature rise. Conversely, a high thermal conductivity substrate (e.g., silicon) allows for greater amounts of heat to be dissipated from the sample, resulting in the need for higher currents to induce behavior similar to that observed in substrates with lower thermal conductivity. This substrate dependence not only reinforces the thermal origin of the second dependency interaction, but also offers a practical route for engineering PTs characteristics in practical devices. 

Beyond these physical implications, our results also carry an experimental advantage. Showing that a single-layer superconducting network system reproduces the full behavior previously associated with ISN structures significantly simplifies the fabrication process of measured samples. Single-network devices require fewer processing steps and reduce variations introduced by inter-layer deposition. In addition, ISN devices inherently suffer from the practical difficulty of ensuring that the two networks remain electrically isolated. Even minor defects or non-uniformities can lead to unintended shorts between the layers. In single-network devices, this issue does not arise at all, eliminating one of the major sources of device failure in ISN fabrication. This makes the devices more reproducible, shortens the time between fabrication iterations, and enables faster exploration of new geometries and substrate conditions. Consequently, the TISS framework can support more efficient experimental studies and open further opportunities for probing these interaction-driven transitions in superconducting systems.

\section{Materials and Methods}

\subsection{Experimental setup and measurements}

In all the experimental measurements, the electric current was driven to each network by a Keithley 2410 sourcemeter, while the voltage across it was measured by a Keithley 2000 multimeter. Sample temperature was controlled and measured by a LakeShore 330 using a $25 \Omega$ heater and a DT-670 thermometer.

For ISN samples, prior to each experiment, we tested and confirmed the absence of short-circuits between the networks by measuring the junction resistance between each pair of cross contacts. Measurements are performed by passing the same current value through both networks simultaneously.

Dynamical measurements were performed after determining the critical current $I_c$ of the N-S PT. We start by setting the current of the system deep inside the N phase. After the resistance stabilizes, we abruptly change the current to a value in the S phase and measure the time $\tau$ it takes for the system to collapse into it. We repeat this process, where in every iteration, the jump of the current gets closer to the critical value until we cross it, and the system does not collapse into the S phase anymore.


The comparison between two different substrates was performed using two TISS samples that were fabricated side by side under identical conditions. Transport measurements were performed on both samples simultaneously.

\subsection{RSJJ model of disordered superconducting networks}
To characterize the N-S PT observed in the experiments, we model the network using a disordered two-dimensional lattice of Resistively Shunted Josephson junctions (RSJJs). In the limit of a large tunneling conductance ($g \gg 1$), isolated networks of RSJJs undergo a continuous N-S PT at low temperatures that are generally independent of the ratio between the Josephson $E_J$ and the Coulomb $E_C$ energies 
\cite{orr1986global,chakravarty1986onset,chakravarty1988quantum}. In this regime, the state of each junction can be characterized by the value of its normal state resistance $R_n(T)$ and by its critical current $I_c(T)$, which generally depends on the ratio between the temperature $T$ of the cryostat and the N-S activation threshold $T_c$. The latter quantities satisfy the Ambegaokar–Baratoff relation \cite{ambegaokar1963tunneling} $I_c(T) R_n = \frac{\pi}{2e} \Delta(T) \tanh(\Delta(T) /2 k_B T)$, where the energy gap, $\Delta(T)$, follows the Bardeen–Cooper–Schrieffer mean-field spectral relation $2\Delta(T) \approx \alpha k_B T_c$ with $\alpha \approx 3.53$, $k_B$ is the Boltzmann constant, and $e$ is the elementary charge. However, the InO networks fabricated in the present work have bulk N-S thresholds large enough to ensure that junctions rarely undergo a metal–insulator transition. In light of this, we consider a model of RSJJ with only three electronic states: superconducting (SC), intermediate (IM), and normal metal (N), defined according to the Josephson I–V characteristic \cite{josephson1962possible}. Hence, the junction’s resistance is defined piecewise as:

\begin{equation}
R_{ij}=\begin{cases}
			R_{\epsilon}, & \text{if $V_{ij} < R_{\epsilon}^n I_{ij}^c(T)$~(SC)}\\
                R_{ij}^n, & \text{if $V_{ij} > R_{ij}^n I_{ij}^c(T)$~(N)}\\
            V_{ij}/I_{ij}^c(T), & \text{otherwise~(IM)}
		 \end{cases}
         \label{eq:JJ_characteristics}
\end{equation}

where $R_{\epsilon}$ is the resistance in the SC state ($R_{\epsilon} = 10^{-5} \Omega$ in simulations), and $V_{ij}$ is the potential drop measured at the ends of the junction. For critical currents, we used a local generalization of the de Gennes relation \cite{de1976relation}: 

\begin{equation}
    I_{ij}^c(T_{ij}) = I_{ij}^c(0)(1 - T_{ij}/ T_{ij}^c)^2
    \label{eq:de-Gennes}
\end{equation}

where $I_{ij}^c(0)$ is the critical current of the junctions at $T = 0$. We control the degree of disorder in the array by considering a quenched normal distribution $\chi_{ij} \in \mathcal{N}(0, \sigma)$, where the variables match the junction's label, the mean is zero, and the variance is $\sigma = 0.1$, serving as a generator for the observables of the other junctions. Generating a unique $\chi_{ij}$ distribution for the network allows us to model the fluctuations in the sample-to-sample disorder level observed in the experiment, which results from the sample fabrication method. In particular, we define $I_{ij}^c(0) = I_0^c(1 + \chi_{ij})$ and $T_{ij}^c = T_c(1 + \chi_{ij})$, where the parameters $I_0^c = 58 \mu A$, $T_c = 2.9K$, and $R_n = 0.5 k\Omega$ were used in simulations.

\subsection{Thermal coupling}
To quantify the self-thermal coupling within the network, we evaluate the power dissipated by each resistor and its influence on the surrounding resistors. The power generated by segment $(i,j)$ is given by the standard electrical power relation:
\begin{equation}
    P_{ij} = \frac{V_{ij}^{2}}{R_{ij}},
    \label{eq:electric_power}
\end{equation}
where $V_{ij}$ and $R_{ij}$ denote the voltage across the resistor and its resistance, respectively.

To determine the thermal contribution of each resistor to all others, we use the heat equation:
\begin{equation}
    \frac{\partial P_{ij}}{\partial t} = D \nabla^{2} P_{ij},
    \label{eq:diffusion}
\end{equation}
where $D$ is the thermal diffusion constant.

The local power $Q_{ij}$ “experienced” by resistor $(i,j)$ is obtained by summing the thermally diffused contributions from all other resistors in the network, using the two-dimensional Green’s function solution of the heat equation:
\begin{equation}
    Q_{ij}^{t}
    = \sum_{(k,l)} P_{ij,kl}^{t}
    = \sum_{(k,l)} \frac{P_{kl}^{t}}{4\pi D t}
        \exp\!\left(-\frac{|i-k|^{2} + |j-l|^{2}}{4Dt}\right),
    \label{eq:diffusion_sol}
\end{equation}
where $|i-k|$ and $|j-l|$ represent the two-dimensional distances between the emitting and receiving resistors. The lattice constant is set to unity. Periodic boundary conditions are imposed to eliminate edge-related effects.

The product $Dt$ effectively characterizes the spatial scale over which heat diffuses and can thus be interpreted as a single diffusion parameter. The analysis assumes a uniform heat distribution consistent with a mean-field approximation. Consequently, the diffusion parameter $Dt$ is chosen to be large compared to the characteristic system size, ensuring that heat is nearly uniformly distributed across all network resistors, as expected from mean-field theory. In the simulations, we used $Dt = 1000 > L = 100$. The diffusion equation, solved using the spectral method, is efficiently evaluated with low numerical cost via fast Fourier transforms (FFTs), making it a computationally advantageous approach for modeling heat dispersion in our network system.

The thermal self-coupling is governed by the properties of the substrate, such that global thermal coupling determines the quasi-static state of the system. The updated global effective temperature at the $t$-th overheating cascade is therefore given iteratively by
\begin{equation}
    \mathbf{T}_{ij}^t = T_0 + \gamma Q_{ij}^t ,
    \label{eq: global_thermal_coupling}
\end{equation}
where $T_0$ is the temperature of the heat bath and $\gamma$ is the heat retention coefficient of the network (the parameter used in simulations). consequently, the heat retention coefficient is inversely proportional to the thermal conductivity of the network substrate. In the simulations, we used $\gamma = 0.1 \times 10^9~\mathrm{W^{-1}\,K}$ for Fig.~\ref{RT}(f) and $\gamma = 2.5 \times 10^9~\mathrm{W^{-1}\,K}$ for Fig.~\ref{RT}(i). For Fig.~\ref{substrate}, we used $\gamma_g = 10 \times 10^9~\mathrm{W^{-1}\,K}$ for the glass substrate and $\gamma_s = 2.5 \times 10^9~\mathrm{W^{-1}\,K}$ for the silicon substrate (the latter having a lower $\gamma$ due to its higher thermal conductivity and more efficient heat dissipation).

\subsection{Thermally coupled Kirchhoff equations}
We developed a model of a thermally self-coupled RSJJ network in which thermal interactions arise from the heat dissipated by individual junctions. Numerical solutions for the order parameter (here, the global sheet resistance $R$) are obtained recursively by allowing the layer to self-interact through its isolated behavior adaptively \cite{ponta2009resistive}. In our model of a self-coupled thermal RSJJ network, this is accomplished by solving the Kirchhoff equations of the array under the adaptive interplay between ~\eqref{eq:JJ_characteristics}, \eqref{eq:de-Gennes}, and \eqref{eq:diffusion_sol}. We consider a two-dimensional lattice of linear size $L$, with its left and right boundaries connected to an external super-node (source) for current injection and to ground, respectively. Each junction follows a Josephson $I$–$V$ characteristic with $R_{ij}$ defined as in~\eqref{eq:JJ_characteristics}, where we take $R_{\epsilon} = 10^{-5}~\Omega$ for the SC state resistance and $R_n = 0.5~\mathrm{k}\Omega$ for the N state resistance. The algorithm is initialized by assigning resistances based on the phase state: when starting from the mutual SC phase, we set $R_{ij,0} = R_{\epsilon}$, whilst for an initial mutual N phase, we set $R_{ij,0} = R_n$.

The algorithm evolves iteratively as follows:\\
\noindent
(i) In the $t$-th stage ($t \geq 1$) of the overheating cascade, the symmetric conductance matrix $\mathbf{G}_t$ is constructed from the junction resistances $R_{ij,t-1}$, with entries
$$
G_{ij}=\begin{cases}
    0, & \text{if $(i,j) \notin E$},\\
    -1/R_{ij}, & \text{if $(i,j) \in E$},\\
    \sum_{k \in \partial i} 1 / R_{ik}, & \text{if $i = j$},
\end{cases}
$$
where $E$ is the set of edges in the array and $\partial i$ is the set of nearest neighbors of node $i$;\\

\noindent
(ii) The potential vector $\mathbf{V}_{t}$ is obtained by numerically solving the Kirchhoff matrix equation
\begin{equation}
    \mathbf{G}_t \cdot \mathbf{V}_{t} = \mathbf{I}_{\mathrm{inj}},
\end{equation}
where ($\cdot$) denotes matrix multiplication and $\mathbf{I}_{\mathrm{inj}}$ is the vector of injected nodal currents, whose entries are zero except for the first component (the super-node) which equals the driving current $I_b$;\\

\noindent
(iii) The heat emitted by each junction is computed through its electrical power, using ~\eqref{eq:electric_power}, with the updated resistance $R_{ij,t}$ and the potential drop $V_{ij,t}$ extracted from $\mathbf{V}_t$;\\

\noindent
(iv) Heat is subsequently thermally diffused across the network. Using the spectral method to solve~\eqref{eq:diffusion}, the contributions are summed via~\eqref{eq:diffusion_sol}, and the effective temperature $\mathbf{T}_t$ is computed using~\eqref{eq: global_thermal_coupling};\\

\noindent
(v) The critical currents $I_{ij}^c(T)$ are updated using~\eqref{eq:de-Gennes}, and the resistive state $R_{ij,t}$ of each junction is determined according to~\eqref{eq:JJ_characteristics};\\

\noindent
(vi) The global resistance of the network is computed as $R_t = \mathbf{V}_t(N) / I_b$.\\

\par
Steps (i)–(vi) are repeated recursively to generate a sequence of potential vectors $\{\mathbf{V}_{1}, \dots, \mathbf{V}_{t}, \dots\}$. Convergence is established once the error,
\begin{equation*}
    \delta \mathbf{V} = \left| 1 - \frac{\mathbf{V}_{t}}{\mathbf{V}_{t+1}} \right|,
\end{equation*}
falls below a numerical threshold $\epsilon_{\min}$. In the simulations, we used $\epsilon_{\min} = 10^{-5}$.

\section*{Data and Software Availability}
Raw data is available upon request from the corresponding author(s). Source code can be freely accessed at the GitHub repository: \\
https://github.com/genutne/From-Interdependent-Networks-to-Two-Interaction-Physical-Systems.\\

\begin{acknowledgments}
S.H. acknowledges the support of the Israel Science Foundation (Grant No. 201/25), the EU H2020 DIT4TRAM, EU H2020 project OMINO (Grant No. 101086321), and the VATAT National Foundation for Climate  and  Power Grid networks and the Israel Ministry of Innovation, Science and Technology (grant number 01017980), for financial support. Y.S., I.V., and A.F. acknowledge support from the Israel Science Foundation (ISF) Grants No. 3053/23 and No. 1499/21. X.W. and B.K. acknowledge the support of the European Research Council grant No. ERC-866236, the ISF grant No. 228/22, and the German Israeli Project Cooperation DIP No. KA 3970/1-1. We thank Ivan Bonnamassa and Nadav Shnerb for useful discussions. 
\end{acknowledgments}

\section*{AUTHORS CONTRIBUTION}
Y.S. and I.V. fabricated the samples. Y.S. performed the experiments. X.W. and B.K. performed the SQUID imaging. N.Y. and B.G. performed the theoretical calculations. A.F. and S.H. initiated and supervised the project. Y.S, N.Y, B.G., A.F., and S.H. wrote the manuscript.

\section*{COMPETING INTERESTS}
The authors declare no competing interests.

\bibliography{LastChance/ForArxivBIB}

\begin{thebibliography}{49}%
\makeatletter
\providecommand \@ifxundefined [1]{%
 \@ifx{#1\undefined}
}%
\providecommand \@ifnum [1]{%
 \ifnum #1\expandafter \@firstoftwo
 \else \expandafter \@secondoftwo
 \fi
}%
\providecommand \@ifx [1]{%
 \ifx #1\expandafter \@firstoftwo
 \else \expandafter \@secondoftwo
 \fi
}%
\providecommand \natexlab [1]{#1}%
\providecommand \enquote  [1]{``#1''}%
\providecommand \bibnamefont  [1]{#1}%
\providecommand \bibfnamefont [1]{#1}%
\providecommand \citenamefont [1]{#1}%
\providecommand \href@noop [0]{\@secondoftwo}%
\providecommand \href [0]{\begingroup \@sanitize@url \@href}%
\providecommand \@href[1]{\@@startlink{#1}\@@href}%
\providecommand \@@href[1]{\endgroup#1\@@endlink}%
\providecommand \@sanitize@url [0]{\catcode `\\12\catcode `\$12\catcode `\&12\catcode `\#12\catcode `\^12\catcode `\_12\catcode `\%12\relax}%
\providecommand \@@startlink[1]{}%
\providecommand \@@endlink[0]{}%
\providecommand \url  [0]{\begingroup\@sanitize@url \@url }%
\providecommand \@url [1]{\endgroup\@href {#1}{\urlprefix }}%
\providecommand \urlprefix  [0]{URL }%
\providecommand \Eprint [0]{\href }%
\providecommand \doibase [0]{https://doi.org/}%
\providecommand \selectlanguage [0]{\@gobble}%
\providecommand \bibinfo  [0]{\@secondoftwo}%
\providecommand \bibfield  [0]{\@secondoftwo}%
\providecommand \translation [1]{[#1]}%
\providecommand \BibitemOpen [0]{}%
\providecommand \bibitemStop [0]{}%
\providecommand \bibitemNoStop [0]{.\EOS\space}%
\providecommand \EOS [0]{\spacefactor3000\relax}%
\providecommand \BibitemShut  [1]{\csname bibitem#1\endcsname}%
\let\auto@bib@innerbib\@empty
\bibitem [{\citenamefont {Binder}(1987)}]{K_Binder_1987}%
  \BibitemOpen
  \bibfield  {author} {\bibinfo {author} {\bibfnamefont {K.}~\bibnamefont {Binder}},\ }\href {https://doi.org/10.1088/0034-4885/50/7/001} {\bibfield  {journal} {\bibinfo  {journal} {Reports on Progress in Physics}\ }\textbf {\bibinfo {volume} {50}},\ \bibinfo {pages} {783} (\bibinfo {year} {1987})}\BibitemShut {NoStop}%
\bibitem [{\citenamefont {Stanley}(1971)}]{stanley1971phase}%
  \BibitemOpen
  \bibfield  {author} {\bibinfo {author} {\bibfnamefont {H.~E.}\ \bibnamefont {Stanley}},\ }\href@noop {} {\emph {\bibinfo {title} {Phase transitions and critical phenomena}}},\ Vol.~\bibinfo {volume} {7}\ (\bibinfo  {publisher} {Clarendon Press, Oxford},\ \bibinfo {year} {1971})\BibitemShut {NoStop}%
\bibitem [{\citenamefont {Domb}(2000)}]{domb2000phase}%
  \BibitemOpen
  \bibfield  {author} {\bibinfo {author} {\bibfnamefont {C.}~\bibnamefont {Domb}},\ }\href@noop {} {\emph {\bibinfo {title} {Phase transitions and critical phenomena}}}\ (\bibinfo  {publisher} {Elsevier},\ \bibinfo {year} {2000})\BibitemShut {NoStop}%
\bibitem [{\citenamefont {Lee}\ \emph {et~al.}(2017)\citenamefont {Lee}, \citenamefont {Choi}, \citenamefont {Kert{\'e}sz},\ and\ \citenamefont {Kahng}}]{lee2017universal}%
  \BibitemOpen
  \bibfield  {author} {\bibinfo {author} {\bibfnamefont {D.}~\bibnamefont {Lee}}, \bibinfo {author} {\bibfnamefont {W.}~\bibnamefont {Choi}}, \bibinfo {author} {\bibfnamefont {J.}~\bibnamefont {Kert{\'e}sz}},\ and\ \bibinfo {author} {\bibfnamefont {B.}~\bibnamefont {Kahng}},\ }\href@noop {} {\bibfield  {journal} {\bibinfo  {journal} {Scientific Reports}\ }\textbf {\bibinfo {volume} {7}} (\bibinfo {year} {2017})}\BibitemShut {NoStop}%
\bibitem [{\citenamefont {Mukamel}(2024)}]{mukamel2024mixed}%
  \BibitemOpen
  \bibfield  {author} {\bibinfo {author} {\bibfnamefont {D.}~\bibnamefont {Mukamel}},\ }in\ \href@noop {} {\emph {\bibinfo {booktitle} {50 Years of the Renormalization Group: Dedicated to the Memory of Michael E Fisher}}}\ (\bibinfo  {publisher} {World Scientific},\ \bibinfo {year} {2024})\ pp.\ \bibinfo {pages} {89--101}\BibitemShut {NoStop}%
\bibitem [{\citenamefont {Boccaletti}\ \emph {et~al.}(2016)\citenamefont {Boccaletti}, \citenamefont {Almendral}, \citenamefont {Guan}, \citenamefont {Leyva}, \citenamefont {Liu}, \citenamefont {Sendi{\~n}a-Nadal}, \citenamefont {Wang},\ and\ \citenamefont {Zou}}]{boccaletti2016explosive}%
  \BibitemOpen
  \bibfield  {author} {\bibinfo {author} {\bibfnamefont {S.}~\bibnamefont {Boccaletti}}, \bibinfo {author} {\bibfnamefont {J.}~\bibnamefont {Almendral}}, \bibinfo {author} {\bibfnamefont {S.}~\bibnamefont {Guan}}, \bibinfo {author} {\bibfnamefont {I.}~\bibnamefont {Leyva}}, \bibinfo {author} {\bibfnamefont {Z.}~\bibnamefont {Liu}}, \bibinfo {author} {\bibfnamefont {I.}~\bibnamefont {Sendi{\~n}a-Nadal}}, \bibinfo {author} {\bibfnamefont {Z.}~\bibnamefont {Wang}},\ and\ \bibinfo {author} {\bibfnamefont {Y.}~\bibnamefont {Zou}},\ }\href@noop {} {\bibfield  {journal} {\bibinfo  {journal} {Physics Reports}\ }\textbf {\bibinfo {volume} {660}},\ \bibinfo {pages} {1} (\bibinfo {year} {2016})}\BibitemShut {NoStop}%
\bibitem [{\citenamefont {Gross}\ \emph {et~al.}(2022)\citenamefont {Gross}, \citenamefont {Bonamassa},\ and\ \citenamefont {Havlin}}]{gross2022fractal}%
  \BibitemOpen
  \bibfield  {author} {\bibinfo {author} {\bibfnamefont {B.}~\bibnamefont {Gross}}, \bibinfo {author} {\bibfnamefont {I.}~\bibnamefont {Bonamassa}},\ and\ \bibinfo {author} {\bibfnamefont {S.}~\bibnamefont {Havlin}},\ }\href@noop {} {\bibfield  {journal} {\bibinfo  {journal} {Physical Review Letters}\ }\textbf {\bibinfo {volume} {129}},\ \bibinfo {pages} {268301} (\bibinfo {year} {2022})}\BibitemShut {NoStop}%
\bibitem [{\citenamefont {Alert}\ \emph {et~al.}(2017)\citenamefont {Alert}, \citenamefont {Tierno},\ and\ \citenamefont {Casademunt}}]{alert2017mixed}%
  \BibitemOpen
  \bibfield  {author} {\bibinfo {author} {\bibfnamefont {R.}~\bibnamefont {Alert}}, \bibinfo {author} {\bibfnamefont {P.}~\bibnamefont {Tierno}},\ and\ \bibinfo {author} {\bibfnamefont {J.}~\bibnamefont {Casademunt}},\ }\href@noop {} {\bibfield  {journal} {\bibinfo  {journal} {Proceedings of the National Academy of Sciences}\ }\textbf {\bibinfo {volume} {114}},\ \bibinfo {pages} {12906} (\bibinfo {year} {2017})}\BibitemShut {NoStop}%
\bibitem [{\citenamefont {Sethna}(2006)}]{sethna2006statistical}%
  \BibitemOpen
  \bibfield  {author} {\bibinfo {author} {\bibfnamefont {J.}~\bibnamefont {Sethna}},\ }\href@noop {} {\emph {\bibinfo {title} {Statistical mechanics: entropy, order parameters, and complexity}}},\ Vol.~\bibinfo {volume} {14}\ (\bibinfo  {publisher} {Oxford University Press},\ \bibinfo {year} {2006})\BibitemShut {NoStop}%
\bibitem [{\citenamefont {Fisher}(1969)}]{fisher1969phase}%
  \BibitemOpen
  \bibfield  {author} {\bibinfo {author} {\bibfnamefont {M.~E.}\ \bibnamefont {Fisher}},\ }in\ \href@noop {} {\emph {\bibinfo {booktitle} {Contemporary Physics: Trieste Symposium 1968. Vol. I. Proceedings of the International Symposium on Contemporary Physics}}}\ (\bibinfo {year} {1969})\BibitemShut {NoStop}%
\bibitem [{\citenamefont {Onsager}(1944)}]{onsager1944crystal}%
  \BibitemOpen
  \bibfield  {author} {\bibinfo {author} {\bibfnamefont {L.}~\bibnamefont {Onsager}},\ }\href@noop {} {\bibfield  {journal} {\bibinfo  {journal} {Physical Review}\ }\textbf {\bibinfo {volume} {65}},\ \bibinfo {pages} {117} (\bibinfo {year} {1944})}\BibitemShut {NoStop}%
\bibitem [{\citenamefont {Kaufman}\ and\ \citenamefont {Onsager}(1949)}]{kaufman1949crystal}%
  \BibitemOpen
  \bibfield  {author} {\bibinfo {author} {\bibfnamefont {B.}~\bibnamefont {Kaufman}}\ and\ \bibinfo {author} {\bibfnamefont {L.}~\bibnamefont {Onsager}},\ }\href@noop {} {\bibfield  {journal} {\bibinfo  {journal} {Physical Review}\ }\textbf {\bibinfo {volume} {76}},\ \bibinfo {pages} {1244} (\bibinfo {year} {1949})}\BibitemShut {NoStop}%
\bibitem [{\citenamefont {Yang}(1952)}]{yang1952spontaneous}%
  \BibitemOpen
  \bibfield  {author} {\bibinfo {author} {\bibfnamefont {C.~N.}\ \bibnamefont {Yang}},\ }\href@noop {} {\bibfield  {journal} {\bibinfo  {journal} {Physical Review}\ }\textbf {\bibinfo {volume} {85}},\ \bibinfo {pages} {808} (\bibinfo {year} {1952})}\BibitemShut {NoStop}%
\bibitem [{\citenamefont {Baxter}(2011)}]{baxter2011onsager}%
  \BibitemOpen
  \bibfield  {author} {\bibinfo {author} {\bibfnamefont {R.}~\bibnamefont {Baxter}},\ }\href@noop {} {\bibfield  {journal} {\bibinfo  {journal} {Journal of Statistical Physics}\ }\textbf {\bibinfo {volume} {145}},\ \bibinfo {pages} {518} (\bibinfo {year} {2011})}\BibitemShut {NoStop}%
\bibitem [{\citenamefont {Kaufman}(1949)}]{kaufman1949crystal2}%
  \BibitemOpen
  \bibfield  {author} {\bibinfo {author} {\bibfnamefont {B.}~\bibnamefont {Kaufman}},\ }\href@noop {} {\bibfield  {journal} {\bibinfo  {journal} {Physical Review}\ }\textbf {\bibinfo {volume} {76}},\ \bibinfo {pages} {1232} (\bibinfo {year} {1949})}\BibitemShut {NoStop}%
\bibitem [{\citenamefont {Cipra}(1987)}]{cipra1987introduction}%
  \BibitemOpen
  \bibfield  {author} {\bibinfo {author} {\bibfnamefont {B.~A.}\ \bibnamefont {Cipra}},\ }\href@noop {} {\bibfield  {journal} {\bibinfo  {journal} {The American Mathematical Monthly}\ }\textbf {\bibinfo {volume} {94}},\ \bibinfo {pages} {937} (\bibinfo {year} {1987})}\BibitemShut {NoStop}%
\bibitem [{\citenamefont {Glauber}(1963)}]{glauber1963time}%
  \BibitemOpen
  \bibfield  {author} {\bibinfo {author} {\bibfnamefont {R.~J.}\ \bibnamefont {Glauber}},\ }\href@noop {} {\bibfield  {journal} {\bibinfo  {journal} {Journal of mathematical physics}\ }\textbf {\bibinfo {volume} {4}},\ \bibinfo {pages} {294} (\bibinfo {year} {1963})}\BibitemShut {NoStop}%
\bibitem [{\citenamefont {Buldyrev}\ \emph {et~al.}(2010)\citenamefont {Buldyrev}, \citenamefont {Parshani}, \citenamefont {Paul}, \citenamefont {Stanley},\ and\ \citenamefont {Havlin}}]{buldyrev-nature2010}%
  \BibitemOpen
  \bibfield  {author} {\bibinfo {author} {\bibfnamefont {S.~V.}\ \bibnamefont {Buldyrev}}, \bibinfo {author} {\bibfnamefont {R.}~\bibnamefont {Parshani}}, \bibinfo {author} {\bibfnamefont {G.}~\bibnamefont {Paul}}, \bibinfo {author} {\bibfnamefont {H.~E.}\ \bibnamefont {Stanley}},\ and\ \bibinfo {author} {\bibfnamefont {S.}~\bibnamefont {Havlin}},\ }\href {https://doi.org/10.1038/nature08932} {\bibfield  {journal} {\bibinfo  {journal} {Nature}\ }\textbf {\bibinfo {volume} {464}},\ \bibinfo {pages} {1025} (\bibinfo {year} {2010})}\BibitemShut {NoStop}%
\bibitem [{\citenamefont {Gao}\ \emph {et~al.}(2011)\citenamefont {Gao}, \citenamefont {Buldyrev}, \citenamefont {Havlin},\ and\ \citenamefont {Stanley}}]{gao-prl2011}%
  \BibitemOpen
  \bibfield  {author} {\bibinfo {author} {\bibfnamefont {J.}~\bibnamefont {Gao}}, \bibinfo {author} {\bibfnamefont {S.~V.}\ \bibnamefont {Buldyrev}}, \bibinfo {author} {\bibfnamefont {S.}~\bibnamefont {Havlin}},\ and\ \bibinfo {author} {\bibfnamefont {H.~E.}\ \bibnamefont {Stanley}},\ }\href {https://doi.org/10.1103/PhysRevLett.107.195701} {\bibfield  {journal} {\bibinfo  {journal} {Phys. Rev. Lett.}\ }\textbf {\bibinfo {volume} {107}},\ \bibinfo {pages} {195701} (\bibinfo {year} {2011})}\BibitemShut {NoStop}%
\bibitem [{\citenamefont {Gao}\ \emph {et~al.}(2012)\citenamefont {Gao}, \citenamefont {Buldyrev}, \citenamefont {Stanley},\ and\ \citenamefont {Havlin}}]{gao-naturephysics2012}%
  \BibitemOpen
  \bibfield  {author} {\bibinfo {author} {\bibfnamefont {J.}~\bibnamefont {Gao}}, \bibinfo {author} {\bibfnamefont {S.~V.}\ \bibnamefont {Buldyrev}}, \bibinfo {author} {\bibfnamefont {H.~E.}\ \bibnamefont {Stanley}},\ and\ \bibinfo {author} {\bibfnamefont {S.}~\bibnamefont {Havlin}},\ }\href {https://doi.org/10.1038/nphys2180} {\bibfield  {journal} {\bibinfo  {journal} {Nature Physics}\ }\textbf {\bibinfo {volume} {8}},\ \bibinfo {pages} {40} (\bibinfo {year} {2012})}\BibitemShut {NoStop}%
\bibitem [{\citenamefont {Bonamassa}\ \emph {et~al.}(2023)\citenamefont {Bonamassa}, \citenamefont {Gross}, \citenamefont {Laav}, \citenamefont {Volotsenko}, \citenamefont {{Frydman, Aviad}},\ and\ \citenamefont {Havlin}}]{bonamassa2023interdependent}%
  \BibitemOpen
  \bibfield  {author} {\bibinfo {author} {\bibfnamefont {I.}~\bibnamefont {Bonamassa}}, \bibinfo {author} {\bibfnamefont {B.}~\bibnamefont {Gross}}, \bibinfo {author} {\bibfnamefont {M.}~\bibnamefont {Laav}}, \bibinfo {author} {\bibfnamefont {I.}~\bibnamefont {Volotsenko}}, \bibinfo {author} {\bibnamefont {{Frydman, Aviad}}},\ and\ \bibinfo {author} {\bibfnamefont {S.}~\bibnamefont {Havlin}},\ }\href@noop {} {\bibfield  {journal} {\bibinfo  {journal} {Nature Physics}\ }\textbf {\bibinfo {volume} {19}},\ \bibinfo {pages} {1163} (\bibinfo {year} {2023})}\BibitemShut {NoStop}%
\bibitem [{\citenamefont {Bashan}\ \emph {et~al.}(2013)\citenamefont {Bashan}, \citenamefont {Berezin}, \citenamefont {Buldyrev},\ and\ \citenamefont {Havlin}}]{Bashan2013}%
  \BibitemOpen
  \bibfield  {author} {\bibinfo {author} {\bibfnamefont {A.}~\bibnamefont {Bashan}}, \bibinfo {author} {\bibfnamefont {Y.}~\bibnamefont {Berezin}}, \bibinfo {author} {\bibfnamefont {S.~V.}\ \bibnamefont {Buldyrev}},\ and\ \bibinfo {author} {\bibfnamefont {S.}~\bibnamefont {Havlin}},\ }\href {https://doi.org/10.1038/nphys2727} {\bibfield  {journal} {\bibinfo  {journal} {Nature Physics}\ }\textbf {\bibinfo {volume} {9}},\ \bibinfo {pages} {667} (\bibinfo {year} {2013})}\BibitemShut {NoStop}%
\bibitem [{\citenamefont {Motter}\ and\ \citenamefont {Lai}(2002)}]{Motter2002}%
  \BibitemOpen
  \bibfield  {author} {\bibinfo {author} {\bibfnamefont {A.~E.}\ \bibnamefont {Motter}}\ and\ \bibinfo {author} {\bibfnamefont {Y.-C.}\ \bibnamefont {Lai}},\ }\href {https://doi.org/10.1103/PhysRevE.66.065102} {\bibfield  {journal} {\bibinfo  {journal} {Physical Review E}\ }\textbf {\bibinfo {volume} {66}},\ \bibinfo {pages} {065102} (\bibinfo {year} {2002})}\BibitemShut {NoStop}%
\bibitem [{\citenamefont {Brummitt}\ \emph {et~al.}(2012)\citenamefont {Brummitt}, \citenamefont {D'Souza},\ and\ \citenamefont {Leicht}}]{Brummitt2012}%
  \BibitemOpen
  \bibfield  {author} {\bibinfo {author} {\bibfnamefont {C.~D.}\ \bibnamefont {Brummitt}}, \bibinfo {author} {\bibfnamefont {R.~M.}\ \bibnamefont {D'Souza}},\ and\ \bibinfo {author} {\bibfnamefont {E.~A.}\ \bibnamefont {Leicht}},\ }\href {https://doi.org/10.1073/pnas.1110586109} {\bibfield  {journal} {\bibinfo  {journal} {Proceedings of the National Academy of Sciences}\ }\textbf {\bibinfo {volume} {109}},\ \bibinfo {pages} {E680} (\bibinfo {year} {2012})}\BibitemShut {NoStop}%
\bibitem [{\citenamefont {Sugiyama}\ \emph {et~al.}(2008)\citenamefont {Sugiyama}, \citenamefont {Fukui}, \citenamefont {Kikuchi}, \citenamefont {Hasebe}, \citenamefont {Nakayama}, \citenamefont {Nishinari}, \citenamefont {ichiro Tadaki},\ and\ \citenamefont {Yukawa}}]{Sugiyama2008}%
  \BibitemOpen
  \bibfield  {author} {\bibinfo {author} {\bibfnamefont {Y.}~\bibnamefont {Sugiyama}}, \bibinfo {author} {\bibfnamefont {M.}~\bibnamefont {Fukui}}, \bibinfo {author} {\bibfnamefont {M.}~\bibnamefont {Kikuchi}}, \bibinfo {author} {\bibfnamefont {K.}~\bibnamefont {Hasebe}}, \bibinfo {author} {\bibfnamefont {A.}~\bibnamefont {Nakayama}}, \bibinfo {author} {\bibfnamefont {K.}~\bibnamefont {Nishinari}}, \bibinfo {author} {\bibfnamefont {S.}~\bibnamefont {ichiro Tadaki}},\ and\ \bibinfo {author} {\bibfnamefont {S.}~\bibnamefont {Yukawa}},\ }\href {https://doi.org/10.1088/1367-2630/10/3/033001} {\bibfield  {journal} {\bibinfo  {journal} {New Journal of Physics}\ }\textbf {\bibinfo {volume} {10}},\ \bibinfo {pages} {033001} (\bibinfo {year} {2008})}\BibitemShut {NoStop}%
\bibitem [{\citenamefont {Geroliminis}\ and\ \citenamefont {Daganzo}(2008)}]{Geroliminis2008}%
  \BibitemOpen
  \bibfield  {author} {\bibinfo {author} {\bibfnamefont {N.}~\bibnamefont {Geroliminis}}\ and\ \bibinfo {author} {\bibfnamefont {C.~F.}\ \bibnamefont {Daganzo}},\ }\href {https://doi.org/10.1016/j.trb.2008.02.002} {\bibfield  {journal} {\bibinfo  {journal} {Transportation Research Part B: Methodological}\ }\textbf {\bibinfo {volume} {42}},\ \bibinfo {pages} {759} (\bibinfo {year} {2008})}\BibitemShut {NoStop}%
\bibitem [{\citenamefont {Cabana}\ and\ \citenamefont {Touboul}(2013)}]{Cabana2013}%
  \BibitemOpen
  \bibfield  {author} {\bibinfo {author} {\bibfnamefont {T.}~\bibnamefont {Cabana}}\ and\ \bibinfo {author} {\bibfnamefont {J.}~\bibnamefont {Touboul}},\ }\href {https://doi.org/10.1007/s10955-013-0801-y} {\bibfield  {journal} {\bibinfo  {journal} {Journal of Statistical Physics}\ }\textbf {\bibinfo {volume} {153}},\ \bibinfo {pages} {211} (\bibinfo {year} {2013})}\BibitemShut {NoStop}%
\bibitem [{\citenamefont {Deco}\ \emph {et~al.}(2011)\citenamefont {Deco}, \citenamefont {Jirsa}, \citenamefont {McIntosh}, \citenamefont {Sporns},\ and\ \citenamefont {K{\"o}tter}}]{Deco2011}%
  \BibitemOpen
  \bibfield  {author} {\bibinfo {author} {\bibfnamefont {G.}~\bibnamefont {Deco}}, \bibinfo {author} {\bibfnamefont {V.}~\bibnamefont {Jirsa}}, \bibinfo {author} {\bibfnamefont {A.~R.}\ \bibnamefont {McIntosh}}, \bibinfo {author} {\bibfnamefont {O.}~\bibnamefont {Sporns}},\ and\ \bibinfo {author} {\bibfnamefont {R.}~\bibnamefont {K{\"o}tter}},\ }\href {https://doi.org/10.1073/pnas.1109563108} {\bibfield  {journal} {\bibinfo  {journal} {Proceedings of the National Academy of Sciences}\ }\textbf {\bibinfo {volume} {108}},\ \bibinfo {pages} {17193} (\bibinfo {year} {2011})}\BibitemShut {NoStop}%
\bibitem [{\citenamefont {Griffa}\ \emph {et~al.}(2019)\citenamefont {Griffa}, \citenamefont {Lichtensteiger},\ and\ \citenamefont {Ravindra}}]{Griffa2019}%
  \BibitemOpen
  \bibfield  {author} {\bibinfo {author} {\bibfnamefont {M.}~\bibnamefont {Griffa}}, \bibinfo {author} {\bibfnamefont {C.}~\bibnamefont {Lichtensteiger}},\ and\ \bibinfo {author} {\bibfnamefont {N.~M.}\ \bibnamefont {Ravindra}},\ }\href {https://doi.org/10.1063/1.5109727} {\bibfield  {journal} {\bibinfo  {journal} {Journal of Applied Physics}\ }\textbf {\bibinfo {volume} {126}},\ \bibinfo {pages} {075104} (\bibinfo {year} {2019})}\BibitemShut {NoStop}%
\bibitem [{\citenamefont {Crespi}\ \emph {et~al.}(2024)\citenamefont {Crespi}, \citenamefont {Zhang},\ and\ \citenamefont {Marques}}]{Crespi2024}%
  \BibitemOpen
  \bibfield  {author} {\bibinfo {author} {\bibfnamefont {V.~H.}\ \bibnamefont {Crespi}}, \bibinfo {author} {\bibfnamefont {Y.}~\bibnamefont {Zhang}},\ and\ \bibinfo {author} {\bibfnamefont {M.~A.~L.}\ \bibnamefont {Marques}},\ }\href {https://doi.org/10.1103/PhysRevB.109.045401} {\bibfield  {journal} {\bibinfo  {journal} {Physical Review B}\ }\textbf {\bibinfo {volume} {109}},\ \bibinfo {pages} {045401} (\bibinfo {year} {2024})}\BibitemShut {NoStop}%
\bibitem [{\citenamefont {Gross}\ \emph {et~al.}(2025)\citenamefont {Gross}, \citenamefont {Volotsenko}, \citenamefont {Sallem}, \citenamefont {Yadid}, \citenamefont {Bonamassa}, \citenamefont {Havlin},\ and\ \citenamefont {Frydman}}]{gross2024microscopic}%
  \BibitemOpen
  \bibfield  {author} {\bibinfo {author} {\bibfnamefont {B.}~\bibnamefont {Gross}}, \bibinfo {author} {\bibfnamefont {I.}~\bibnamefont {Volotsenko}}, \bibinfo {author} {\bibfnamefont {Y.}~\bibnamefont {Sallem}}, \bibinfo {author} {\bibfnamefont {N.}~\bibnamefont {Yadid}}, \bibinfo {author} {\bibfnamefont {I.}~\bibnamefont {Bonamassa}}, \bibinfo {author} {\bibfnamefont {S.}~\bibnamefont {Havlin}},\ and\ \bibinfo {author} {\bibfnamefont {A.}~\bibnamefont {Frydman}},\ }\href {https://doi.org/10.1038/s41467-025-61127-z} {\bibfield  {journal} {\bibinfo  {journal} {Nature Communications}\ }\textbf {\bibinfo {volume} {16}},\ \bibinfo {pages} {5869} (\bibinfo {year} {2025})}\BibitemShut {NoStop}%
\bibitem [{\citenamefont {Parshani}\ \emph {et~al.}(2010)\citenamefont {Parshani}, \citenamefont {Buldyrev},\ and\ \citenamefont {Havlin}}]{parshani2010}%
  \BibitemOpen
  \bibfield  {author} {\bibinfo {author} {\bibfnamefont {R.}~\bibnamefont {Parshani}}, \bibinfo {author} {\bibfnamefont {S.~V.}\ \bibnamefont {Buldyrev}},\ and\ \bibinfo {author} {\bibfnamefont {S.}~\bibnamefont {Havlin}},\ }\href {https://doi.org/10.1103/PhysRevLett.105.048701} {\bibfield  {journal} {\bibinfo  {journal} {Phys. Rev. Lett.}\ }\textbf {\bibinfo {volume} {105}},\ \bibinfo {pages} {048701} (\bibinfo {year} {2010})}\BibitemShut {NoStop}%
\bibitem [{\citenamefont {Gross}\ \emph {et~al.}(2024)\citenamefont {Gross}, \citenamefont {Bonamassa},\ and\ \citenamefont {Havlin}}]{gross2024microscopic2}%
  \BibitemOpen
  \bibfield  {author} {\bibinfo {author} {\bibfnamefont {B.}~\bibnamefont {Gross}}, \bibinfo {author} {\bibfnamefont {I.}~\bibnamefont {Bonamassa}},\ and\ \bibinfo {author} {\bibfnamefont {S.}~\bibnamefont {Havlin}},\ }\href@noop {} {\bibfield  {journal} {\bibinfo  {journal} {Physical Review Letters}\ }\textbf {\bibinfo {volume} {132}},\ \bibinfo {pages} {227401} (\bibinfo {year} {2024})}\BibitemShut {NoStop}%
\bibitem [{\citenamefont {Kirtley}\ \emph {et~al.}(1995)\citenamefont {Kirtley}, \citenamefont {Ketchen}, \citenamefont {Stawiasz}, \citenamefont {Sun}, \citenamefont {Gallagher}, \citenamefont {Blanton},\ and\ \citenamefont {Wind}}]{Kirtley1995APL}%
  \BibitemOpen
  \bibfield  {author} {\bibinfo {author} {\bibfnamefont {J.~R.}\ \bibnamefont {Kirtley}}, \bibinfo {author} {\bibfnamefont {M.~B.}\ \bibnamefont {Ketchen}}, \bibinfo {author} {\bibfnamefont {K.~G.}\ \bibnamefont {Stawiasz}}, \bibinfo {author} {\bibfnamefont {J.~Z.}\ \bibnamefont {Sun}}, \bibinfo {author} {\bibfnamefont {W.~J.}\ \bibnamefont {Gallagher}}, \bibinfo {author} {\bibfnamefont {S.~H.}\ \bibnamefont {Blanton}},\ and\ \bibinfo {author} {\bibfnamefont {S.~J.}\ \bibnamefont {Wind}},\ }\href {https://doi.org/10.1063/1.113838} {\bibfield  {journal} {\bibinfo  {journal} {Applied Physics Letters}\ }\textbf {\bibinfo {volume} {66}},\ \bibinfo {pages} {1138} (\bibinfo {year} {1995})}\BibitemShut {NoStop}%
\bibitem [{\citenamefont {Wang}\ \emph {et~al.}(2022)\citenamefont {Wang}, \citenamefont {Laav}, \citenamefont {Volotsenko}, \citenamefont {Frydman},\ and\ \citenamefont {Kalisky}}]{Xi2022PRA}%
  \BibitemOpen
  \bibfield  {author} {\bibinfo {author} {\bibfnamefont {X.}~\bibnamefont {Wang}}, \bibinfo {author} {\bibfnamefont {M.}~\bibnamefont {Laav}}, \bibinfo {author} {\bibfnamefont {I.}~\bibnamefont {Volotsenko}}, \bibinfo {author} {\bibfnamefont {A.}~\bibnamefont {Frydman}},\ and\ \bibinfo {author} {\bibfnamefont {B.}~\bibnamefont {Kalisky}},\ }\href {https://doi.org/10.1103/PhysRevApplied.17.024073} {\bibfield  {journal} {\bibinfo  {journal} {Phys. Rev. Appl.}\ }\textbf {\bibinfo {volume} {17}},\ \bibinfo {pages} {024073} (\bibinfo {year} {2022})}\BibitemShut {NoStop}%
\bibitem [{\citenamefont {Roth}\ \emph {et~al.}(1989)\citenamefont {Roth}, \citenamefont {Sepulveda},\ and\ \citenamefont {Wikswo}}]{Roth1989APL}%
  \BibitemOpen
  \bibfield  {author} {\bibinfo {author} {\bibfnamefont {B.~J.}\ \bibnamefont {Roth}}, \bibinfo {author} {\bibfnamefont {N.~G.}\ \bibnamefont {Sepulveda}},\ and\ \bibinfo {author} {\bibfnamefont {J.}~\bibnamefont {Wikswo}, \bibfnamefont {John~P.}},\ }\href {https://doi.org/10.1063/1.342549} {\bibfield  {journal} {\bibinfo  {journal} {Journal of Applied Physics}\ }\textbf {\bibinfo {volume} {65}},\ \bibinfo {pages} {361} (\bibinfo {year} {1989})}\BibitemShut {NoStop}%
\bibitem [{\citenamefont {Zhou}\ \emph {et~al.}(2014)\citenamefont {Zhou}, \citenamefont {Bashan}, \citenamefont {Cohen}, \citenamefont {Berezin}, \citenamefont {Shnerb},\ and\ \citenamefont {Havlin}}]{dong-pre2014}%
  \BibitemOpen
  \bibfield  {author} {\bibinfo {author} {\bibfnamefont {D.}~\bibnamefont {Zhou}}, \bibinfo {author} {\bibfnamefont {A.}~\bibnamefont {Bashan}}, \bibinfo {author} {\bibfnamefont {R.}~\bibnamefont {Cohen}}, \bibinfo {author} {\bibfnamefont {Y.}~\bibnamefont {Berezin}}, \bibinfo {author} {\bibfnamefont {N.}~\bibnamefont {Shnerb}},\ and\ \bibinfo {author} {\bibfnamefont {S.}~\bibnamefont {Havlin}},\ }\href {https://doi.org/10.1103/PhysRevE.90.012803} {\bibfield  {journal} {\bibinfo  {journal} {Phys. Rev. E}\ }\textbf {\bibinfo {volume} {90}},\ \bibinfo {pages} {012803} (\bibinfo {year} {2014})}\BibitemShut {NoStop}%
\bibitem [{\citenamefont {Baxter}\ \emph {et~al.}(2015)\citenamefont {Baxter}, \citenamefont {Dorogovtsev}, \citenamefont {Lee}, \citenamefont {Mendes},\ and\ \citenamefont {Goltsev}}]{baxter2015critical}%
  \BibitemOpen
  \bibfield  {author} {\bibinfo {author} {\bibfnamefont {G.}~\bibnamefont {Baxter}}, \bibinfo {author} {\bibfnamefont {S.}~\bibnamefont {Dorogovtsev}}, \bibinfo {author} {\bibfnamefont {K.-E.}\ \bibnamefont {Lee}}, \bibinfo {author} {\bibfnamefont {J.}~\bibnamefont {Mendes}},\ and\ \bibinfo {author} {\bibfnamefont {A.}~\bibnamefont {Goltsev}},\ }\href@noop {} {\bibfield  {journal} {\bibinfo  {journal} {Physical Review X}\ }\textbf {\bibinfo {volume} {5}},\ \bibinfo {pages} {031017} (\bibinfo {year} {2015})}\BibitemShut {NoStop}%
\bibitem [{\citenamefont {Hohenberg}\ and\ \citenamefont {Halperin}(1977)}]{halperin1977}%
  \BibitemOpen
  \bibfield  {author} {\bibinfo {author} {\bibfnamefont {P.~C.}\ \bibnamefont {Hohenberg}}\ and\ \bibinfo {author} {\bibfnamefont {B.~I.}\ \bibnamefont {Halperin}},\ }\href {https://doi.org/10.1103/RevModPhys.49.435} {\bibfield  {journal} {\bibinfo  {journal} {Rev. Mod. Phys.}\ }\textbf {\bibinfo {volume} {49}},\ \bibinfo {pages} {435} (\bibinfo {year} {1977})}\BibitemShut {NoStop}%
\bibitem [{\citenamefont {Thompson}\ and\ \citenamefont {Younglove}(1961)}]{THOMPSON1961146}%
  \BibitemOpen
  \bibfield  {author} {\bibinfo {author} {\bibfnamefont {J.}~\bibnamefont {Thompson}}\ and\ \bibinfo {author} {\bibfnamefont {B.}~\bibnamefont {Younglove}},\ }\href {https://doi.org/https://doi.org/10.1016/0022-3697(61)90146-9} {\bibfield  {journal} {\bibinfo  {journal} {Journal of Physics and Chemistry of Solids}\ }\textbf {\bibinfo {volume} {20}},\ \bibinfo {pages} {146} (\bibinfo {year} {1961})}\BibitemShut {NoStop}%
\bibitem [{\citenamefont {Zaitlin}\ and\ \citenamefont {Anderson}(1974)}]{PhysRevLett.33.1158}%
  \BibitemOpen
  \bibfield  {author} {\bibinfo {author} {\bibfnamefont {M.~P.}\ \bibnamefont {Zaitlin}}\ and\ \bibinfo {author} {\bibfnamefont {A.~C.}\ \bibnamefont {Anderson}},\ }\href {https://doi.org/10.1103/PhysRevLett.33.1158} {\bibfield  {journal} {\bibinfo  {journal} {Phys. Rev. Lett.}\ }\textbf {\bibinfo {volume} {33}},\ \bibinfo {pages} {1158} (\bibinfo {year} {1974})}\BibitemShut {NoStop}%
\bibitem [{\citenamefont {Bonamassa}\ \emph {et~al.}(2025)\citenamefont {Bonamassa}, \citenamefont {Gross}, \citenamefont {Kert{\'e}sz},\ and\ \citenamefont {Havlin}}]{bonamassa2025hybrid}%
  \BibitemOpen
  \bibfield  {author} {\bibinfo {author} {\bibfnamefont {I.}~\bibnamefont {Bonamassa}}, \bibinfo {author} {\bibfnamefont {B.}~\bibnamefont {Gross}}, \bibinfo {author} {\bibfnamefont {J.}~\bibnamefont {Kert{\'e}sz}},\ and\ \bibinfo {author} {\bibfnamefont {S.}~\bibnamefont {Havlin}},\ }\href@noop {} {\bibfield  {journal} {\bibinfo  {journal} {Nature Communications}\ }\textbf {\bibinfo {volume} {16}},\ \bibinfo {pages} {1415} (\bibinfo {year} {2025})}\BibitemShut {NoStop}%
\bibitem [{\citenamefont {Orr}\ \emph {et~al.}(1986)\citenamefont {Orr}, \citenamefont {Jaeger}, \citenamefont {Goldman},\ and\ \citenamefont {Kuper}}]{orr1986global}%
  \BibitemOpen
  \bibfield  {author} {\bibinfo {author} {\bibfnamefont {B.}~\bibnamefont {Orr}}, \bibinfo {author} {\bibfnamefont {H.}~\bibnamefont {Jaeger}}, \bibinfo {author} {\bibfnamefont {A.}~\bibnamefont {Goldman}},\ and\ \bibinfo {author} {\bibfnamefont {C.}~\bibnamefont {Kuper}},\ }\href@noop {} {\bibfield  {journal} {\bibinfo  {journal} {Physical review letters}\ }\textbf {\bibinfo {volume} {56}},\ \bibinfo {pages} {378} (\bibinfo {year} {1986})}\BibitemShut {NoStop}%
\bibitem [{\citenamefont {Chakravarty}\ \emph {et~al.}(1986)\citenamefont {Chakravarty}, \citenamefont {Ingold}, \citenamefont {Kivelson},\ and\ \citenamefont {Luther}}]{chakravarty1986onset}%
  \BibitemOpen
  \bibfield  {author} {\bibinfo {author} {\bibfnamefont {S.}~\bibnamefont {Chakravarty}}, \bibinfo {author} {\bibfnamefont {G.-L.}\ \bibnamefont {Ingold}}, \bibinfo {author} {\bibfnamefont {S.}~\bibnamefont {Kivelson}},\ and\ \bibinfo {author} {\bibfnamefont {A.}~\bibnamefont {Luther}},\ }\href@noop {} {\bibfield  {journal} {\bibinfo  {journal} {Physical review letters}\ }\textbf {\bibinfo {volume} {56}},\ \bibinfo {pages} {2303} (\bibinfo {year} {1986})}\BibitemShut {NoStop}%
\bibitem [{\citenamefont {Chakravarty}\ \emph {et~al.}(1988)\citenamefont {Chakravarty}, \citenamefont {Ingold}, \citenamefont {Kivelson},\ and\ \citenamefont {Zimanyi}}]{chakravarty1988quantum}%
  \BibitemOpen
  \bibfield  {author} {\bibinfo {author} {\bibfnamefont {S.}~\bibnamefont {Chakravarty}}, \bibinfo {author} {\bibfnamefont {G.-L.}\ \bibnamefont {Ingold}}, \bibinfo {author} {\bibfnamefont {S.}~\bibnamefont {Kivelson}},\ and\ \bibinfo {author} {\bibfnamefont {G.}~\bibnamefont {Zimanyi}},\ }\href@noop {} {\bibfield  {journal} {\bibinfo  {journal} {Physical Review B}\ }\textbf {\bibinfo {volume} {37}},\ \bibinfo {pages} {3283} (\bibinfo {year} {1988})}\BibitemShut {NoStop}%
\bibitem [{\citenamefont {Ambegaokar}\ and\ \citenamefont {Baratoff}(1963)}]{ambegaokar1963tunneling}%
  \BibitemOpen
  \bibfield  {author} {\bibinfo {author} {\bibfnamefont {V.}~\bibnamefont {Ambegaokar}}\ and\ \bibinfo {author} {\bibfnamefont {A.}~\bibnamefont {Baratoff}},\ }\href@noop {} {\bibfield  {journal} {\bibinfo  {journal} {Physical Review Letters}\ }\textbf {\bibinfo {volume} {10}},\ \bibinfo {pages} {486} (\bibinfo {year} {1963})}\BibitemShut {NoStop}%
\bibitem [{\citenamefont {Josephson}(1962)}]{josephson1962possible}%
  \BibitemOpen
  \bibfield  {author} {\bibinfo {author} {\bibfnamefont {B.~D.}\ \bibnamefont {Josephson}},\ }\href@noop {} {\bibfield  {journal} {\bibinfo  {journal} {Physics letters}\ }\textbf {\bibinfo {volume} {1}},\ \bibinfo {pages} {251} (\bibinfo {year} {1962})}\BibitemShut {NoStop}%
\bibitem [{\citenamefont {De~Gennes}(1976)}]{de1976relation}%
  \BibitemOpen
  \bibfield  {author} {\bibinfo {author} {\bibfnamefont {P.-G.}\ \bibnamefont {De~Gennes}},\ }\href@noop {} {\bibfield  {journal} {\bibinfo  {journal} {Journal de Physique Lettres}\ }\textbf {\bibinfo {volume} {37}},\ \bibinfo {pages} {1} (\bibinfo {year} {1976})}\BibitemShut {NoStop}%
\bibitem [{\citenamefont {Ponta}\ \emph {et~al.}(2009)\citenamefont {Ponta}, \citenamefont {Carbone}, \citenamefont {Gilli},\ and\ \citenamefont {Mazzetti}}]{ponta2009resistive}%
  \BibitemOpen
  \bibfield  {author} {\bibinfo {author} {\bibfnamefont {L.}~\bibnamefont {Ponta}}, \bibinfo {author} {\bibfnamefont {A.}~\bibnamefont {Carbone}}, \bibinfo {author} {\bibfnamefont {M.}~\bibnamefont {Gilli}},\ and\ \bibinfo {author} {\bibfnamefont {P.}~\bibnamefont {Mazzetti}},\ }\href@noop {} {\bibfield  {journal} {\bibinfo  {journal} {Physical Review B}\ }\textbf {\bibinfo {volume} {79}},\ \bibinfo {pages} {134513} (\bibinfo {year} {2009})}\BibitemShut {NoStop}%
\end{thebibliography}%

\clearpage
\newpage
\onecolumngrid
\section*{Supplementary Information}

\setcounter{figure}{0} 
\renewcommand{\thefigure}{S\arabic{figure}}
\renewcommand{\figurename}{FIG.}

\begin{figure}[h]
    \centering
    \includegraphics[width=\textwidth]{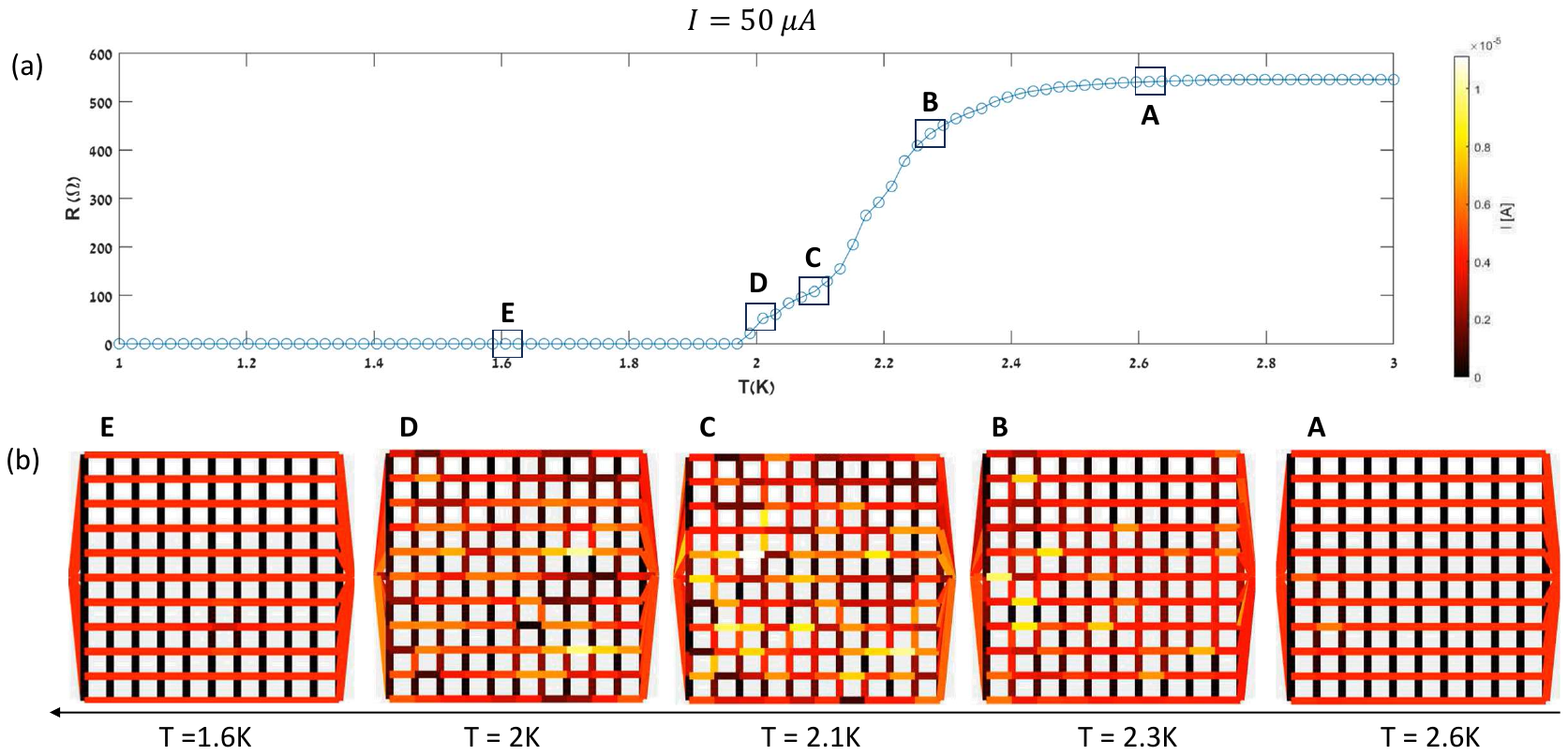}
    \caption{\textbf{Network current-flow map across a continuous N-S phase transition.}
    Numerical simulations for cooling the system under a fixed current of $I = 50\,\mu\mathrm{A}$.
    The network size is $L = 11$. Heat spreads uniformly according to the mean-field approximation, where the temporal diffusion parameter satisfies $D t = L^2 = 121$. 
    We used $\gamma = 1.452 \times 10^{6}\,\mathrm{W^{-1}\,K}$. \textbf{(a)} Network resistance as a function of temperature during cooling from $T = 3\,\mathrm{K}$ down to $T = 1\,\mathrm{K}$, showing a continuous second-order N-S transition. \textbf{(b)} Representative maps of the network’s current-flow-intensity configuration at five characteristic temperatures, indicated by the square markers in panel~(a). The imposed current flows from the left boundary to the right, and the total resistance is measured between the corresponding injection and extraction nodes. During the phase transition (B-D), the currents spontaneously percolate through high-density current paths.}

\end{figure}

\begin{figure}[h]
    \centering
    \includegraphics[width=\textwidth]{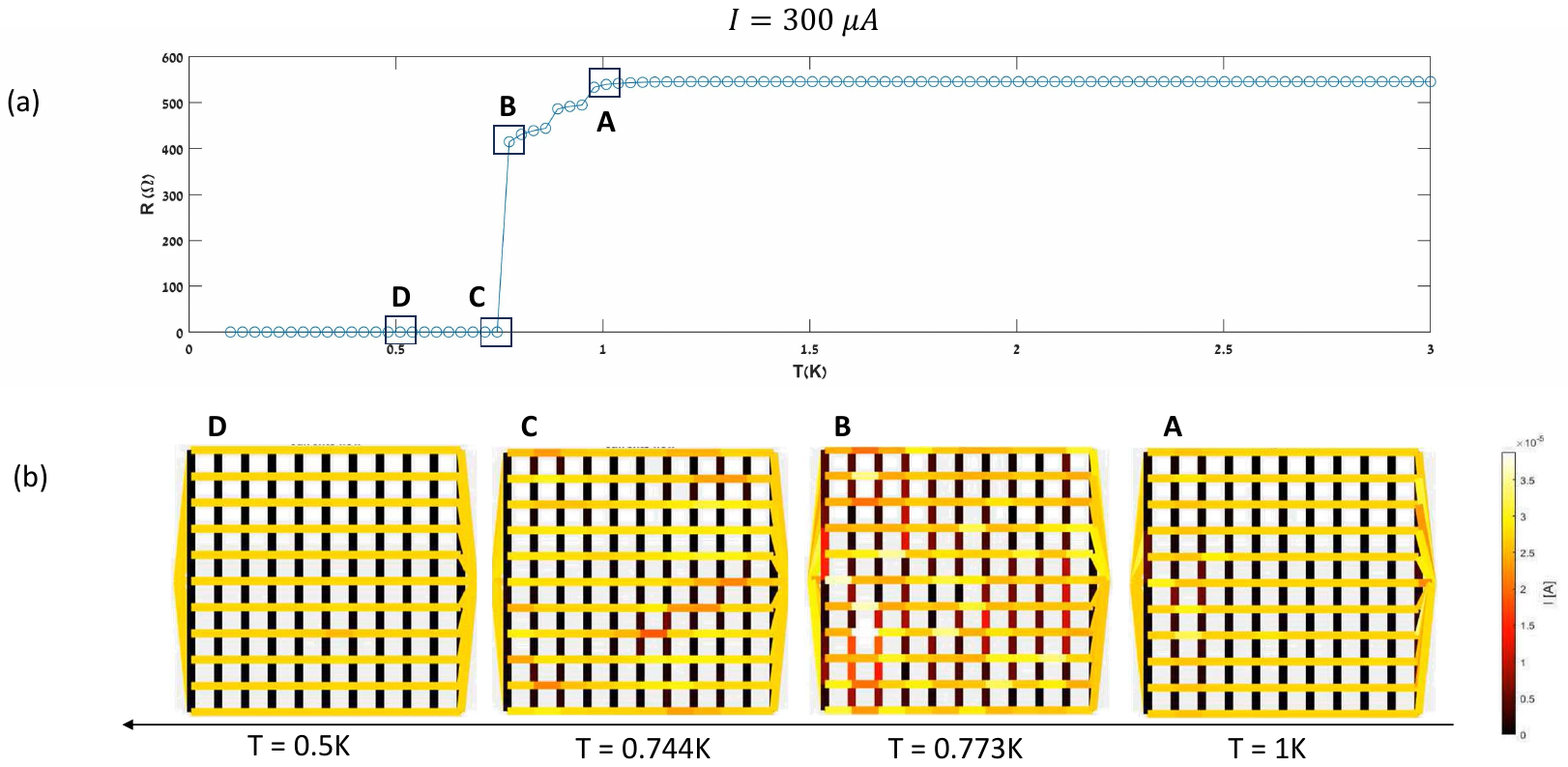}
    \caption{\textbf{Network current-flow map across an abrupt N-S phase transition.}
    Numerical simulations for cooling the system under a fixed current of $I = 300\,\mu\mathrm{A}$. 
    The network size is $L = 11$. Heat spreads uniformly according to the mean-field approximation, where the temporal diffusion parameter satisfies $D t = L^2 = 121$. 
    We used $\gamma = 1.452 \times 10^{6}\,\mathrm{W^{-1}\,K}$.
    \textbf{(a)} Network resistance as a function of temperature during cooling from $T = 3\,\mathrm{K}$ down to $T = 0.1\,\mathrm{K}$, showing an abrupt N-S phase transition. \textbf{(b)} Representative maps of the network’s current-flow-intensity configuration at four characteristic temperatures, indicated by the square markers in panel~(a). The imposed current flows from the left boundary to the right, and the total resistance is measured between the corresponding injection and extraction nodes. In contrast to the continuous second-order transition (Fig.~S1\textbf{B-D}), no percolation structure is observed in the high-density current paths in the vicinity of the transition point (B, C).}

\end{figure}

\newpage

\begin{figure}[h]
    \centering
    \includegraphics[width=\textwidth]{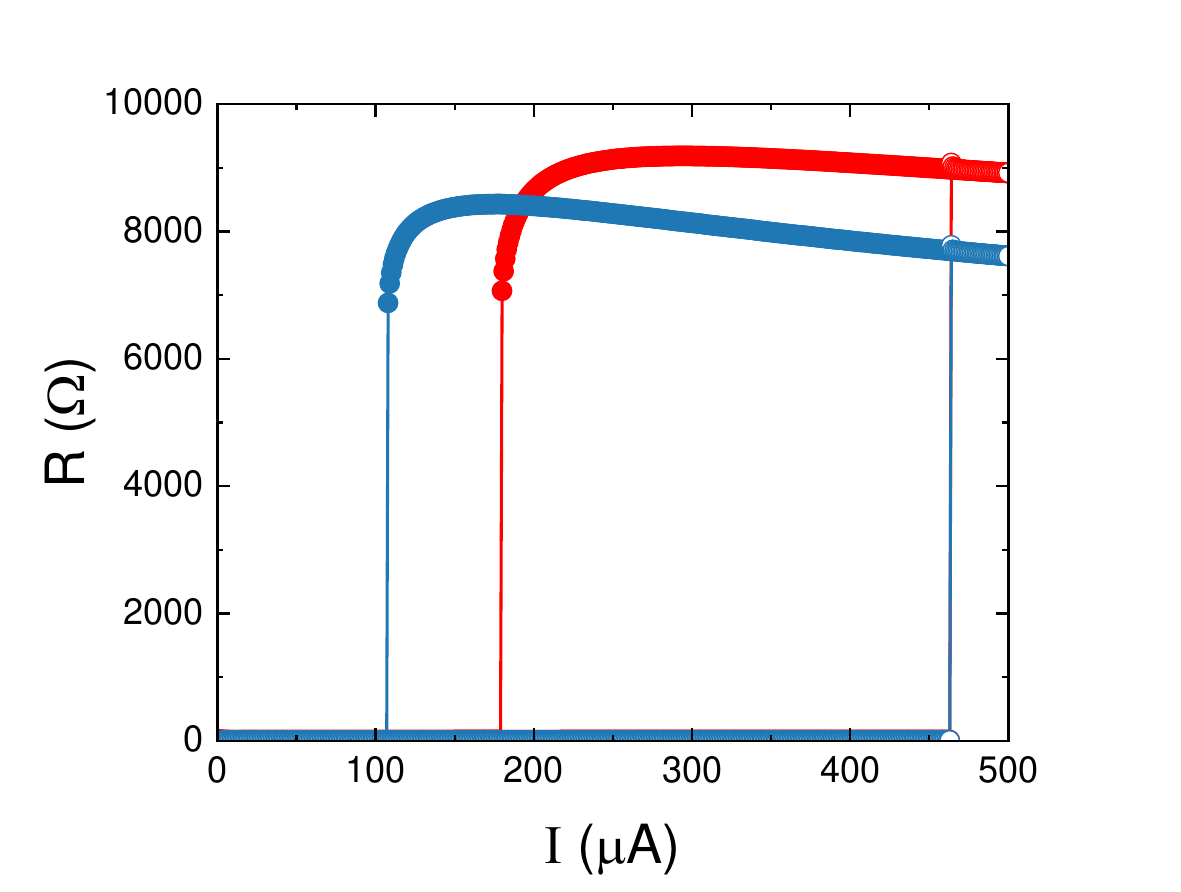}
    \caption{\textbf{RI substrate dependence curves -} Experimental resistance versus current curves for the same samples of Fig. 3. Here, red is for the network on silicon and blue is for the network on glass. Full symbols represent decreasing current curves (N-S transition), and empty symbols represent increasing current curves (S-N transition). Throughout all measurements, the temperature is kept constant at $T=1.8$K.}
\end{figure}

\end{document}